\documentclass[fleqn,usenatbib]{mnras}

\usepackage{newtxtext,newtxmath}

\usepackage[T1]{fontenc}
\usepackage{ae,aecompl}

\usepackage{graphicx}	
\usepackage{amsmath}	

\usepackage{amssymb}	
\usepackage{orcidlink}
\usepackage{anyfontsize}
\usepackage{soul}

\newcommand{\tess}{{\it TESS}}

\newcommand{\gaia}{{\it Gaia}}

\newcommand{\harps}{{\it HARPS}}
\newcommand{\coralie}{{\it CORALIE}}

\newcommand{\NGTS}{NGTS}
\newcommand{\TESS}{{\it TESS}}

\newcommand{\kms}{km\,s$^{-1}$}

\newcommand{\mpl}{\mbox{M\textsubscript{p}}}
\newcommand{\rpl}{\mbox{R\textsubscript{p}}}
\newcommand{\mstar}{\mbox{M$_{\star}$}}
\newcommand{\rstar}{\mbox{R$_{\star}$}}

\newcommand{\mjup}{\mbox{M\textsubscript{J}}}
\newcommand{\rjup}{\mbox{R\textsubscript{J}}}
\newcommand{\msun}{\mbox{M$_{\odot}$}}
\newcommand{\rsun}{\mbox{R$_{\odot}$}}

\newcommand{\rearth}{R$_{\oplus}$}

\newcommand{\gcc}{g\,cm$^{-3}$}

\newcommand{\alles}{\texttt{allesfitter}}

\newcommand{\tiara}{\texttt{TIaRA}}
\newcommand{\bsproc}{\texttt{bsproc}}

\newcommand{\ldtk}{\textsc{ldtk}}

\newcommand{\teff}{T\textsubscript{eff}}

\newcommand{\teq}{$T_{\rm eq}$}
\newcommand{\logg}{$\log g$}

\newcommand{\vsini}{$v \sin i_\star$}

\newcommand{\feh}{[Fe/H]}

\newcommand{\TICID}{453147896}
\newcommand{\starname}{NGTS-39}
\newcommand{\planetname}{NGTS-39\,b}
\newcommand{\GaiaID}{3156473342454773248}
\newcommand{\twoMASSID}{J07194529+0956286}

\newcommand{\rahms}{$07:19:45.29$}
\newcommand{\dechms}{$+09:56:28.55$}

\newcommand{\pmra}{$-10.0423\pm 0.0207$}
\newcommand{\pmdec}{$-2.9010\pm0.01538$}
\newcommand{\parallax}{$3.584\pm0.0.018$}

\newcommand{\Vmag}{$11.516\pm0.048$}
\newcommand{\Bmag}{$12.139\pm0.044$}
\newcommand{\GaiaGmag}{$11.4221\pm0.0004$}
\newcommand{\GaiaBPmag}{$11.7190\pm0.0005$}
\newcommand{\GaiaRPmag}{$10.9675\pm0.0004$}
\newcommand{\TESSmag}{$11.025\pm0.006$}
\newcommand{\TESSmagshort}{$11.02$}
\newcommand{\Jmag}{$10.475\pm0.024$}
\newcommand{\Hmag}{$10.187\pm 0.023$}
\newcommand{\Kmag}{$10.197\pm0.026$}
\newcommand{\WOnemag}{$10.137\pm0.023$}
\newcommand{\WTwomag}{$10.178\pm0.02$}
\newcommand{\WThreemag}{$10.199\pm0.076$}

\newcommand{\hostteff}{$6053_{-30}^{+67}$}
\newcommand{\hostlogg}{$4.419\pm0.042$}
\newcommand{\hostfeh}{$0.196\pm0.023$}
\newcommand{\hostvsini}{$4.0\pm0.5$}

\newcommand{\hostrad}{$1.16\pm0.01$}
\newcommand{\hostmass}{$1.16\pm0.05$}
\newcommand{\hostage}{$2.2\pm0.8$}
\newcommand{\brr}{$0.09651\pm0.00060$} 
\newcommand{\brsuma}{$0.01921_{-0.00029}^{+0.00031}$} 
\newcommand{\bcosi}{$0.01812\pm0.00071$} 
\newcommand{\bepoch}{$2460157.03626\pm0.00055$} 
\newcommand{\bperiod}{$58.204720_{-0.000038}^{+0.000042}$}
\newcommand{\bfc}{$-0.090\pm0.033$} 
\newcommand{\bfs}{$0.614\pm0.016$} 
\newcommand{\bK}{$0.0755\pm0.0022$} 
\newcommand{\hostldcqoneNGTSone}{$0.3397\pm0.010$} 
\newcommand{\hostldcqtwoNGTSone}{$0.284\pm0.070$} 
\newcommand{\hostldcqoneNGTSsix}{$0.3397\pm0.010$} 
\newcommand{\hostldcqtwoNGTSsix}{$0.284\pm0.070$} 
\newcommand{\hostldcqoneTESS}{$0.2871\pm0.0084$} 
\newcommand{\hostldcqtwoTESS}{$0.339\pm0.074$} 
\newcommand{\hostldcqonesgone}{$0.2891\pm0.0083$} 
\newcommand{\hostldcqtwosgone}{$0.363\pm0.069$} 
\newcommand{\baselineslopervHARPS}{$-0.01776\pm0.001855$} 
\newcommand{\baselineoffsetrvHARPS}{$42.0189\pm0.0032$} 
\newcommand{\baselineoffsetrvCORALIEone}{$42.1741\pm0.0056$} 
\newcommand{\baselineoffsetrvCORALIEtwo}{$42.1462\pm0.0036$} 

\newcommand{\bRstarovera}{$0.01751_{-0.00026}^{+0.00028}$} 
\newcommand{\bRcompanionRjup}{$1.088\pm0.012$} 
\newcommand{\baAU}{$0.3077\pm0.0055$} 
\newcommand{\be}{$0.386\pm0.019$} 
\newcommand{\bw}{$98.3\pm3.1$} 
\newcommand{\periastron}{$0.189\pm0.007$}
\newcommand{\apastron}{$0.426\pm0.010$}
\newcommand{\bMcompanionMjup}{$1.467\pm0.081$} 
\newcommand{\bbtra}{$0.639_{-0.023}^{+0.020}$} 
\newcommand{\bTtratot}{$4.639\pm0.035$} 
\newcommand{\bTtrafull}{$3.325\pm0.046$} 
\newcommand{\bdensity}{$1.411\pm0.092$} 
\newcommand{\bTeq}{$519_{-5}^{+6}$} 
\newcommand{\bTeqPeri}{$662_{-17}^{+18}$} 
\newcommand{\bTeqAp}{$441_{-7}^{+8}$} 
\newcommand{\bdepthtrdilTESS}{$0.0118_{-0.0084}^{+0.014}$} 

\usepackage{color}
\usepackage{subcaption}

\usepackage{gensymb}
\usepackage{multicol}
\usepackage{booktabs,caption}
\usepackage[flushleft]{threeparttable}
\usepackage{pdflscape}

\title[NGTS-39\,b: a 58-day warm Jupiter]{NGTS-39\,b: A 58\,d transiting warm Jupiter in an eccentric orbit}

\author[Ioannis Apergis et al.]{\parbox{\textwidth}{\Large
Ioannis Apergis$^{1,2,3}$\thanks{E-mail: Ioannis.Apergis@warwick.ac.uk;
D.Bayliss@warwick.ac.uk; ulmer-moll@strw.leidenuniv.nl}\orcidlink{0009-0004-7473-4573},
Daniel Bayliss$^{1,2}$\footnotemark[1]\orcidlink{0000-0001-6023-1335},
Sol\`ene Ulmer-Moll$^{4}$\footnotemark[1]\orcidlink{0000-0003-2417-7006},
Samuel~Gill$^{1,2}$\orcidlink{0000-0002-4259-0155},
Toby~Rodel$^{5}$\orcidlink{0009-0009-2175-72841},
Matthew~Battley$^{6}$\orcidlink{0000-0002-1357-9774},
Paul~Benni$^{7}$\orcidlink{0000-0001-6981-8722},
Allyson~Bieryla$^{20}$\orcidlink{0000-0001-6637-5401},
James~A.~Blake$^{1,2}$\orcidlink{0000-0002-5903-2387},
Andrea~Bonfanti$^{8}$,
François~Bouchy$^{9}$\orcidlink{0000-0002-7613-393X},
Edward~M.~Bryant$^{1,2}$\orcidlink{0000-0001-7904-4441},
Matthew~R.~Burleigh$^{10}$\orcidlink{0000-0003-0684-7803},
Samuel~J.~Carlier$^{10}$,
Sarah~L.~Casewell$^{10}$,
Hritam~Chakraborty$^{9}$\orcidlink{0000-0002-5177-1898},
Alastair~B.~Claringbold$^{1,2}$\orcidlink{0000-0003-1309-5558},
Karen A. Collins$^{20}$\orcidlink{0000-0001-6588-9574},
Benjamin~D.~R.~Davies$^{1,2}$\orcidlink{0009-0000-5659-9006},
Xavier~Dumusque$^{9}$\orcidlink{0000-0002-9332-2011},
Troy~A.~Edkins$^{10}$\orcidlink{0009-0003-8688-7836},
Fintan~Eeles-Nolle$^{1,2}$\orcidlink{0009-0009-6207-3217},
Jo~Ann~Egger$^{11}$\orcidlink{0000-0003-1628-4231},
Jorge~Fernández~Fernández$^{1,2}$\orcidlink{0000-0002-1416-2188},
Marcelo~Aron~Fetzner~Keniger$^{1,2}$\orcidlink{0009-0005-2761-9190},
Pedro~Figueira$^{12}$\orcidlink{0000-0001-8504-283X},
Michael~R.~Goad$^{10}$,
George~Harvey$^{10}$,
Faith~Hawthorn$^{13,1}$\orcidlink{0000-0002-8675-182X},
Melissa~J.~Hobson$^{9}$\orcidlink{0000-0002-5945-7975},
Mathilde~Houelle$^{9}$\orcidlink{0009-0001-7217-4099},
Giovanni~Isopi$^{14}$,
Timour~Jestin$^{15}$\orcidlink{0009-0000-5585-7915},
Alicia~Kendall$^{10}$,
Monika~Lendl$^{9}$\orcidlink{0000-0001-9699-1459},
Daniel~Lewis$^{10}$,
Isobel~Lockley$^{1,2}$\orcidlink{0009-0003-0928-3588},
Franco~Mallia$^{14}$,
James~McCormac$^{1,2,3}$\orcidlink{0000-0003-1631-4170},
Morgan~A.~Mitchell$^{1,2}$\orcidlink{0009-0004-6130-7775},
Lucile~Mignon$^{9,27}$\orcidlink{0000-0002-5407-3905},
Hugh~Osborn$^{25}$\orcidlink{0000-0002-4047-4724},
Angelica~Psaridi$^{16,17}$\orcidlink{0000-0002-4797-2419},
Alex~Romanec$^{1,2}$\orcidlink{0009-0009-1748-3428},
Suman~Saha$^{21,22}$\orcidlink{0000-0001-8018-0264},
Amber~Sedgley$^{1,2}$\orcidlink{0009-0001-7824-1715},
Sérgio~Sousa$^{18,19}$,
Neil~Thomas$^{26}$\orcidlink{0000-0002-2146-3894},
Stéphane~Udry$^{9}$\orcidlink{0000-0001-7576-6236},
Christopher~Watson$^{5}$\orcidlink{0000-0002-9718-3266},
Richard~G.~West$^{1,2}$\orcidlink{0000-0001-6604-5533},
Thomas~G.~Wilson$^{1,2}$\orcidlink{0000-0001-8749-1962},
Peter~J.~Wheatley$^{1,2}$\orcidlink{0000-0003-1452-2240},
Jamie~T.~Williams$^{1,2}$\orcidlink{0009-0007-8709-9689},
Aldo~Zapparata$^{14}$\orcidlink{0000-0002-9428-1573},
Krzysztof~Sz.~Zieliński$^{23,24}$\orcidlink{0009-0001-0389-8907}
}
\vspace{0.2cm}
\\
Authors' institutions are listed at the end of the paper.
}
\date{Accepted XXX. Received YYY; in original form ZZZ}

\pubyear{2026}

\begin{document}
\label{firstpage}
\pagerange{\pageref{firstpage}--\pageref{lastpage}}
\maketitle

\begin{abstract}
We report the discovery and characterisation of \planetname\ (TIC\,\TICID\,b), a warm Jupiter transiting a Sun-like star on a 58.2\,day, eccentric ($e$ = \be) orbit. \planetname\ was first identified from a \tess\ single transit event, and subsequently confirmed with NGTS photometry and radial velocity measurements from \coralie\ and \harps.  The host star is a bright ($T_{\mathrm{mag}}$ = \TESSmagshort) F9 dwarf with effective temperature of \teff = \hostteff\,K. \planetname\ is a Jupiter-sized gas giant with a radius of \bRcompanionRjup\,\rjup\ and a mass of \bMcompanionMjup\,\mjup. Its equilibrium temperature is \bTeq\,K, placing it between short-period hot Jupiters and cold, Jupiter-like giants. The high orbital eccentricity and intermediate equilibrium temperature of \planetname\ make it a valuable test case for formation and migration models, particularly in the poorly sampled regime of long-period gas giants. The RV data show a linear trend of $\dot{\gamma} = -17.75$\,m\,s$^{-1}$\,yr$^{-1}$, which indicates the presence of an outer companion. The discovery of \planetname\ contributes to the small but growing population of transiting warm Jupiters with P$>$50 days orbiting bright stars. 
\end{abstract}



\section{Introduction}
\label{sec:intro}
Long-period transiting exoplanets, typically with orbital periods of several tens to hundreds of days, occupy a unique regime in the study of planetary systems \citep{Uehara2016,ForMorHogAgoSch16}. Unlike short-period hot Jupiters, these planets spend most of their orbits far from their host stars, making transits rare and challenging to detect. Their characterisation often requires long-term photometric monitoring to capture multiple transits and confirm orbital periods. 

Space-based missions such as the Transiting Exoplanet Survey Satellite \citep[\tess;][]{Ricker2015TESS}
have substantially expanded the photometric characterisation of Jupiter-like exoplanets, including both newly identified candidates and previously known systems. Owing to their large radii, high geometric transit probabilities, and frequent transits, the most readily detected members of this population are hot Jupiters. While hot Jupiters provide valuable information on planetary formation, evolution, and atmospheric physics \citep{2016ApJ...816...21D,2020MNRAS.498..680G}, they represent a highly irradiated and dynamically altered population. Their short orbital periods ($<$10 days) place them in extreme environments dominated by strong stellar irradiation and tidal interactions, which complicates the interpretation of their formation pathways and atmospheric properties \citep{Baraffeandchabrier2010, Laughlin2018, Sarkis2021}.

In contrast, warm and cold Jupiter-like planets with orbital periods ranging from tens to hundreds of days offer a much clearer window into planetary system architecture and dynamical history \citep{Petrovich_2016, Hamers_2017, Anderson_2020, Morgan_2025}. These planets reside in environments where their atmospheres are less strongly altered by stellar irradiation, allowing more direct probes of their primordial composition \citep{Madhusudhan_2019, Komacek_2020, Fortney2020}. Their wider orbits also make them excellent tracers of migration processes: measurements of their masses, eccentricities, and orbital obliquities from ground-based spectroscopic follow-up can provide further evidence for disc-driven migration and high-eccentricity pathways. Moreover, warm Jupiters often coexist with additional planets, making them key to understanding the architecture and long-term evolution of planetary systems \citep{Huang_2016}.

Detecting long-period transiting planets is inherently challenging. A single \tess\ sector provides only $\sim$27 days of continuous coverage, and most fields are not revisited until two years later. Consequently, long-period planets may produce only a single detectable transit (monotransit) in the \tess\ observing windows \citep{Cooke2018, Cooke2019, Cooke2021, Villanueva2019, Rodel2024}, resulting in unknown orbital period solutions.   

Ground-based facilities such as the Next Generation Transit Survey \citep[\NGTS;][]{Wheatley2018NGTS} have therefore developed dedicated monotransit campaigns to find and observe these \tess\ candidates \citep{Gill2020NGTSMonos, 2024MNRAS.533..109G}. These campaigns follow-up the targets in a blind survey, enabling the recovery of additional transits in order to determine the true orbital period of the systems.

These efforts allow us to discover exoplanets with longer orbital periods and cooler temperatures than those of typical short-period transiting systems. Such longer-period giant planets provide an important comparison sample because they are less affected by intense stellar irradiation, radius inflation, and tidal evolution than hot Jupiters. Their measured radii and bulk densities can therefore help test whether the diversity in heavy-element enrichment and internal structure seen among giant planets depends on orbital separation and irradiation environment\citep{Dalba2022TOI2180LongEccentricPHTESS,UlmerMoll2022NGTS20b_TOI5153bMonos}.

In the last few years, photometric and RV follow-up efforts of \tess\ monotransit candidate led to the discovery and characterisation of transiting long-period exoplanets \citep[e.g.][]{Battley2024NGTS30bMono, 2024MNRAS.533..109G, 2025arXiv250915424U}. In this work, we report the discovery and detailed characterisation of \planetname, a long-period giant initially identified in the \tess\ photometry during the Year 3 extended mission. To constrain its orbital period, we initiated a follow-up campaign with \NGTS\ targeting this monotransit candidate, which successfully observed a second transit three years after the initial \tess\ monotransit event. Subsequently \tess\ observed a third transit  in the \tess\ Year 6 extended mission which was consistent with a 58.2\,d orbital period. The period was further refined using additional \NGTS\ observations and data from the \tess\ Follow-up Observing Program Sub Group 1 team \citep[TFOP SG1;][]{Collins2018TFOP}. In parallel, high precision radial velocity (RV) measurements from \harps\ \citep{Pepe2002HARPS} and \coralie\ \citep{Queloz2001CORALIE} confirmed the orbital and measure the planetary mass, hence establishing the planetary nature of \planetname.

We structure this paper as follows. In Section~\ref{sec:photometry}, we present the \tess\ and \NGTS\ observations of \starname. Section~\ref{sec:rv} describes the RV monitoring conducted with \harps\ and \coralie. In Section~\ref{sec:Results and Analysis}, we perform a joint modelling of the photometric and spectroscopic datasets to derive the system parameters. We discuss our findings in Section~\ref{sec:discussion}, and summarise our conclusions in Section~\ref{sec:conclusion}.

\section{Photometry}
\label{sec:photometry}

\subsection{\tess\ Photometry}
\label{sec:tess_phot}

The \tess\ mission is a NASA space-based observatory designed to obtain high-precision, wide-field photometry for the detection of transiting exoplanets around bright, nearby stars \citep[][]{Ricker2015TESS}. \tess\ is equipped with four identical frame-transfer CCD cameras based on the CCID-80 devices developed by MIT/Lincoln Laboratory. These back-illuminated, deep-depleted sensors provide high sensitivity at red optical wavelengths. Each camera comprises a $2048 \times 2048$ array of active imaging pixels, with an equal number of storage pixels, and a pixel pitch of $15 \times 15$ micron. The plate scale projected on sky is approximately $21\arcsec\,\mathrm{pixel}^{-1}$. Each camera covers a field of view of $24^\circ \times 24^\circ$, resulting in a combined instantaneous field of view of $24^\circ \times 96^\circ$. The cameras operate over a broad red-optical bandpass spanning approximately 600-1000\,nm and observe the sky in sectors of approximately 27 days. During its prime mission, \tess\ first observed the southern ecliptic hemisphere in Year 1 and then moved to the northern ecliptic hemisphere in Year 2, thereby completing a  nearly all-sky transit survey. This strategy provided near-continuous coverage close to the ecliptic poles, where adjacent sectors overlap, while stars nearer the ecliptic plane were typically observed for only a single sector.

\begin{figure}
	\includegraphics[width=\columnwidth]{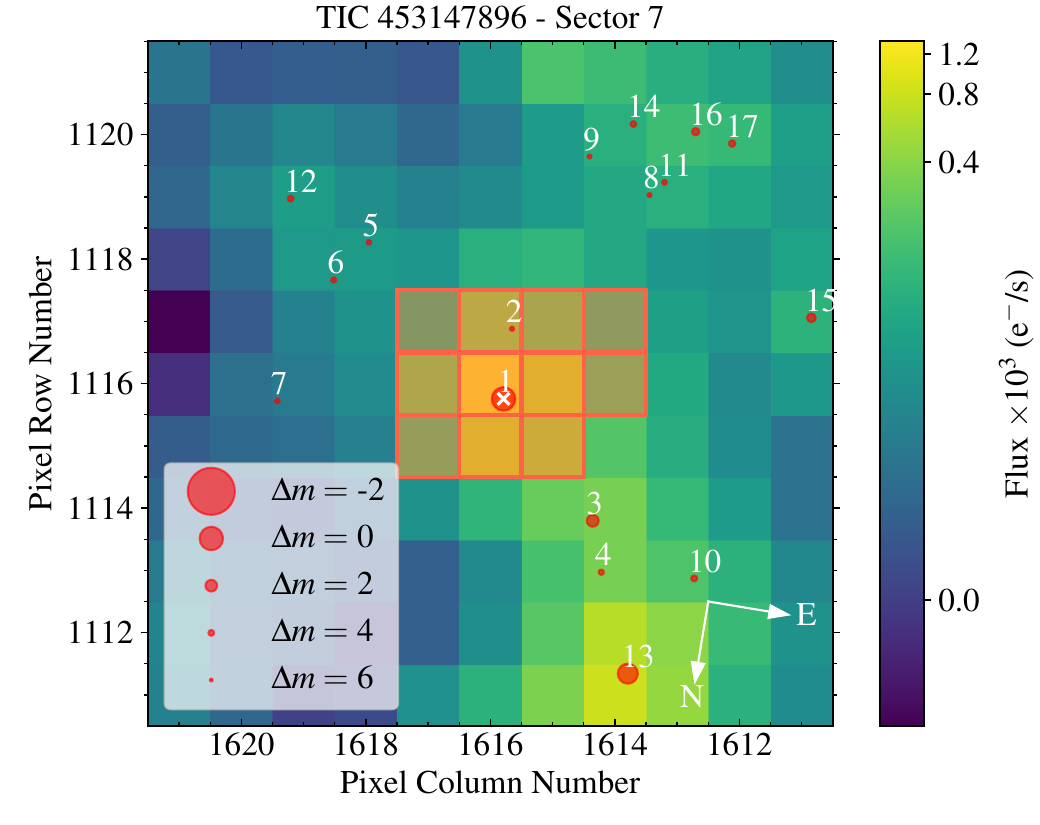}
    \caption{\tess\ Sector 7 Full-Frame Image cutout (11$\times$11 pixels) of \starname\ and the surrounding region, generated with \texttt{tpfplotter} \citep{Aller2020TPFPlotter}. \starname\ is shown at the centre, labelled "1" and marked with the white x. The only additional source falling within our adopted aperture is approximately 6 magnitudes fainter than the target, and therefore its flux contribution is expected to be negligible. No significant dilution is therefore expected in the \tess\ photometry.}
    \label{fig:ind_lcs}
\end{figure}

\begin{figure}
	\includegraphics[width=\columnwidth]{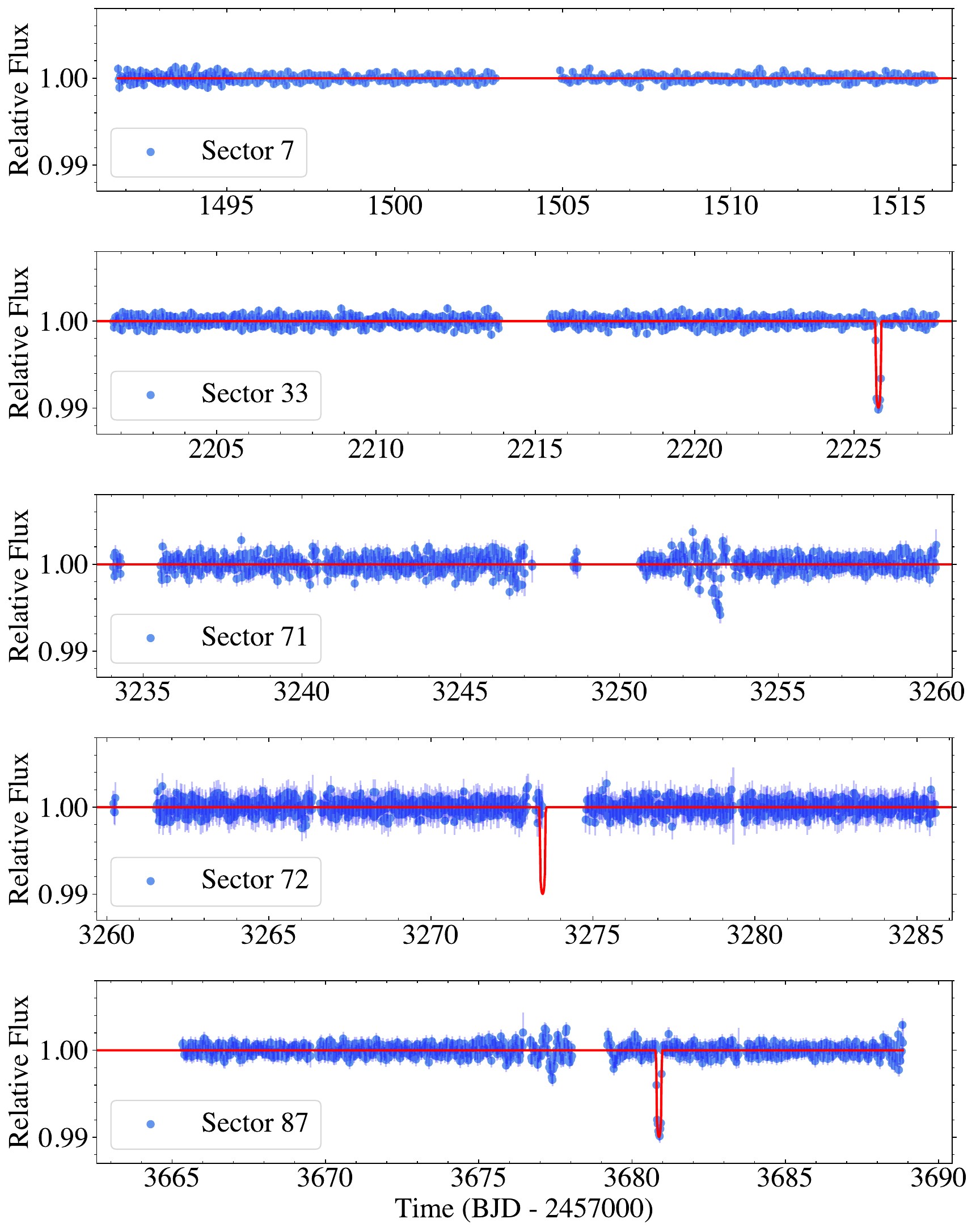}
    \caption{\tess\ Sector detrended and normalised lightcurves zoomed-in for \starname\ from \tess\ full-frame images pipeline. The transit events are marked with red. Transits are detected in Sector 33 and Sector 87. The transit event in Sector 72 was just missed due to the orbit gap in the \tess\ data. All the data has been binned to 30-minutes for clarity and consistency.}
    \label{fig:tess_sectors}
\end{figure}

\starname\ (TIC-\TICID) is a relatively bright ($T_{\mathrm{mag}}$ = \TESSmagshort) F9 dwarf star. The full set of stellar properties for \starname\ are set out in \autoref{tab:host_properties}. For our analysis of \starname, we generate custom light curves directly from the \tess\ Science Processing Operations Centre \citep[SPOC;][]{Jenkins2016TESSSPOC} Full-Frame images. This was done as the SPOC and Quick Look Pipeline \citep[QLP;][]{Huang2020QLPdoi} lightcurves for this star were not always available when we needed them for our analysis. It also allowed us to tailor the photometric aperture and background pixel mask to suit \starname\ in each TESS sector. We use the \texttt{lightkurve} package \citep{Lightkurve2018lightkurve} to download the \TESS\ target pixel files for \starname\ filtered using the SPOC quality flag $\mathrm{QUALITY} = 0$ \citep{Jenkins2016TESSSPOC, TESSSPOC2020}. A pixel-level aperture was constructed as a two-tier mask: background pixels were defined as the 30\,\% lowest-flux pixels in a reference cadence, while target pixels were selected using a threshold mask applied to the summed flux across cadences.  An example of our target pixel mask from Sector 7 is shown in \autoref{fig:ind_lcs}. The target flux was computed by summing all pixels in the target aperture and subtracting the median background flux per pixel multiplied by the number of target pixels. The flux uncertainties were estimated using the median absolute deviation of the simple aperture photometry \texttt{SAP} flux, providing robust per-cadence error estimates. To remove long-term trends while preserving transit signals, the lightcurve was split into segments at gaps exceeding 2.4\,h, corresponding to spacecraft data downloads and SPOC-flagged regions. Each segment was normalised individually and subsequently smoothed using an iterative Savitzky–Golay filter (five iterations with a 3$\sigma$ clipping threshold) and a 48\,h window size, which efficiently removes stellar activity while setting an upper limit on the duration of detectable transit events \citep{Savitzky_1964, Hattori_2022}. We present all of the \tess\ photometry in \autoref{fig:tess_sectors}.

In total, \starname\ has been observed by \tess\ in Sectors 7, 33, 71, 72, and 87 (see \autoref{fig:tess_sectors}). Full details of each Sector of \tess\ observations are set out in \autoref{tab:phot_summary}.

\begin{table}
    \centering
    \caption{\starname\ Stellar parameters}
    \label{tab:host_properties}
    \begin{tabular}{ccc}
         \toprule
         Property & Value & Source \\
         \hline
         \hline
         Spectral type & F9V & \cite{PecautMamajek2013ColorsTemps} \\
         \hline
         \multicolumn{3}{c}{\textit{Identifiers}} \\ 
         \hline
         2MASS ID & \twoMASSID & 2MASS  \\
         \gaia\ Source ID&  \GaiaID & \gaia\ DR3  \\
         TIC ID&   \TICID & TIC 8  \\
         NGTS ID&   \starname & This Work \\
         \hline
         \multicolumn{3}{c}{\textit{Coordinates}} \\ 
         \hline
         RA (hh:mm:ss.ss)&  \rahms & \gaia\ DR3  \\
         DEC (dd:mm:ss.ss)&  \dechms & \gaia\ DR3  \\
         \hline
         \multicolumn{3}{c}{\textit{Proper motion, parallax and systemic velocity}} \\ 
         \hline
         $\mu_\text{RA}$ (mas y$^{-1}$)&  \pmra & \gaia\ DR3  \\
         $\mu_\text{DEC}$ (mas y$^{-1}$)&  \pmdec & \gaia\ DR3  \\
         Parallax (mas)&  \parallax & \gaia\ DR3  \\
         ${\gamma}$\,(m\,s$^{-1}$) & $ 42113.05 \pm 2.44$ & This Work \\
         $\dot{\gamma}$\,(m\,s$^{-1}$\,yr$^{-1}$) & $-17.75 \pm 1.86$ & This Work \\
         \hline
         \multicolumn{3}{c}{\textit{Magnitudes}} \\ 
         \hline
         V (mag) & \Vmag & APASS DR10  \\
         B (mag) & \Bmag & APASS DR10  \\
         G (mag) & \GaiaGmag & \gaia\ DR3  \\
         BP (mag) & \GaiaBPmag & \gaia\ DR3  \\
         RP (mag) & \GaiaRPmag & \gaia\ DR3  \\
         \tess\ (mag) & \TESSmag & TIC 8  \\
         J (mag) & \Jmag & 2MASS  \\
         H (mag) & \Hmag & 2MASS  \\
         K (mag) & \Kmag & 2MASS  \\
         W1 (mag) & \WOnemag & WISE  \\
         W2 (mag) & \WTwomag & WISE  \\
         W3 (mag) & \WThreemag & WISE  \\
         \hline
         \multicolumn{3}{c}{\textit{Spectral Parameters}}\\
         \hline
         \vsini\ (\kms) & \hostvsini & This Work \\
         \teff\ (K) & \hostteff & This Work  \\
         \feh\ (`dex') & \hostfeh & This Work  \\
         \logg\ ($\log(\text{cgs})$) & \hostlogg & This Work \\
         \hline
         \multicolumn{3}{c}{\textit{Derived parameters}} \\ 
         \hline
         \rstar\ (\rsun) & \hostrad & This Work  \\
         \mstar\ (\msun) & \hostmass & This Work  \\
         Age (Gyr) & \hostage & This Work  \\
         \bottomrule
    \end{tabular}
\end{table}

\starname\ was first identified as a monotransit candidate by the NGTS Monotransit working group following the procedure set out in \cite{Hawthorn_2024}.  A transit of duration 4.6 hours and depth 1\,\% was detected in the \tess\ Sector 33 lightcurve centered at BJD\,2459225.7607 (2021-01-11 06:15 UTC) (see \autoref{fig:tess_sectors}). Based on this monotransit event, \starname\ was scheduled for NGTS follow-up and a second transit was observed (see \autoref{sec:ngts_phot}). \starname\ was observed again by \tess\ in Sectors 71 and 72, although no transits were detected during these Sectors.  A subsequent transit was detected in Sector 87, with a depth and duration consistent with the earlier transits detected in Sector 33 and by NGTS.


We note that, based on the orbital period fitted to the \tess\ data, a transit is expected to occur in Sector 72 (see \autoref{fig:tess_sectors}), however we could not retrieve the transit. We investigated the availability of alternative data products such as the Quick Look Pipeline \citep[QLP; ][]{Huang2020QLPdoi}. Upon inspection, the QLP light curve is significantly affected by increased scatter similarly to our custom lightcurve product, likely driven by strong stray light. We measured the average sky background during the transit window to be 35 times higher than the average background before the transit. This introduces substantial systematics that degrade the photometric precision. Crucially, the predicted transit occurs immediately prior to a data gap, in a region where the light curve quality is already compromised. As a result, the transit signal cannot be reliably identified or extracted. We also note that the Sector 71 light curve shows a flux decrease near 3253\,(BJD $-$ 2457000), which could potentially resemble a transit-like event. However this event is most likely a spacecraft systematic, and is immediately prior to a data downlink. The event occurs during a section of poor pointing performance and is common to a large number of stars as evidenced in the data release notes for Sector~71.\footnote{\url{https://archive.stsci.edu/missions/tess/doc/tess_drn/tess_sector_71_drn100_v01.pdf}}

\subsection{NGTS Photometry}
\label{sec:ngts_phot}
\NGTS\ \citep{Wheatley2018NGTS} is an exoplanet-hunting facility located at ESO’s Paranal Observatory in Chile. It comprises twelve robotic Newtonian telescopes, each with a 20\,cm primary mirror and a $\sim$8\,deg$^2$ field of view. This field of view makes NGTS well suited for high precision photometry of bright stars where a wide field of view is critical for monitoring comparison stars needed for relative photometry.

Each telescope is equipped with a deep-depleted CCD camera featuring 13.5 microns pixel pitch and a 2048 $\times$ 2048 usable sensor with enhanced sensitivity at red wavelengths. Given the NGTS telescope and camera configuration, the plate scale projected on sky is approximately $5\arcsec\,\mathrm{pixel}^{-1}$. NGTS uses a custom wide-band filter tailored to the red-sensitive response (520-890\,nm) of these detectors. The excellent photometric conditions at Paranal further enhance NGTS’s capability for precise photometric observations. Multiple telescopes can observe the same target simultaneously, allowing independent measurements to be combined and improving the overall photometric precision \citep{2020MNRAS.494.5872B}. For bright stars, \NGTS\ observations are effectively scintillation-limited without introducing additional correlated noise \citep{OBrien2022Scint}.

Following the monotransit event identified in the \tess\ Sector 33 lightcurve, \starname\ was targeted as part of the \NGTS\ single-transit program. \NGTS\ observations started in 2023/09/04. The initial monitoring was carried out with a single NGTS telescope using 10\,s exposures between 2023/09/04 and 2024/02/04, with observations limited to airmass $<$ 2. In total the \NGTS\ campaign for \starname\ lasted for 104 observing nights. 

Raw images are astrometrically calibrated using a world coordinate system (WCS) solution via \textit{astrometry.net} \citep{astrometry}, incorporating field centre, pixel scale, and a low-order polynomial. To improve accuracy near image edges, we cross-match sources with TIC8 \citep{STScI2018TIC}, selecting stars brighter than $T<16$\,mag, and iteratively refine the fit until sub-pixel precision is reached. Frames lacking a reliable WCS solution are excluded. Stars for photometry are also selected from TIC8 with $T<16$\,mag, avoiding blended sources (neighbours within 6 pixels and $\Delta T<2.5$\,mag) and edge regions (for details see \citet{Apergis_2026}). Daily dark and bias frames are obtained, but no flat-field correction is applied, as autoguiding keeps stars on consistent pixels \citep{DONUTS,Wheatley2018NGTS}. As in \citet{Apergis_2026}, we found that omitting flat-field correction improves the RMS scatter by approximately 0.29\,\% for a typical stellar sample. We therefore consider it safer to omit flat fielding than to apply an imperfect correction. In addition, because the telescopes are very fast (f/2.8), dust on the CCD window is highly out of focus and does not produce strong donut like features. The sealed space between the telescope and CCD window also minimises external dust contamination.

Aperture photometry is performed using \textsc{sep} \citep{sex,sex2016} with circular apertures optimised per camera. Background maps are constructed and subtracted prior to flux extraction. Relative photometry corrects for atmospheric and instrumental trends using comparison stars. Stars are iteratively filtered based on RMS deviations from magnitude-binned trends, removing variable stars. The remaining stars form a master reference light curve used to correct the target star, and the optimal aperture is chosen to minimise the RMS scatter (see details in \citet{2020MNRAS.494.5872B} and the pipeline in  \url{https://github.com/NGTS/bsproc}). An example of the \NGTS\ photometry output is set out in \autoref{tab:ngtsphot}. 

\begin{figure*}
	\includegraphics[width=\textwidth]{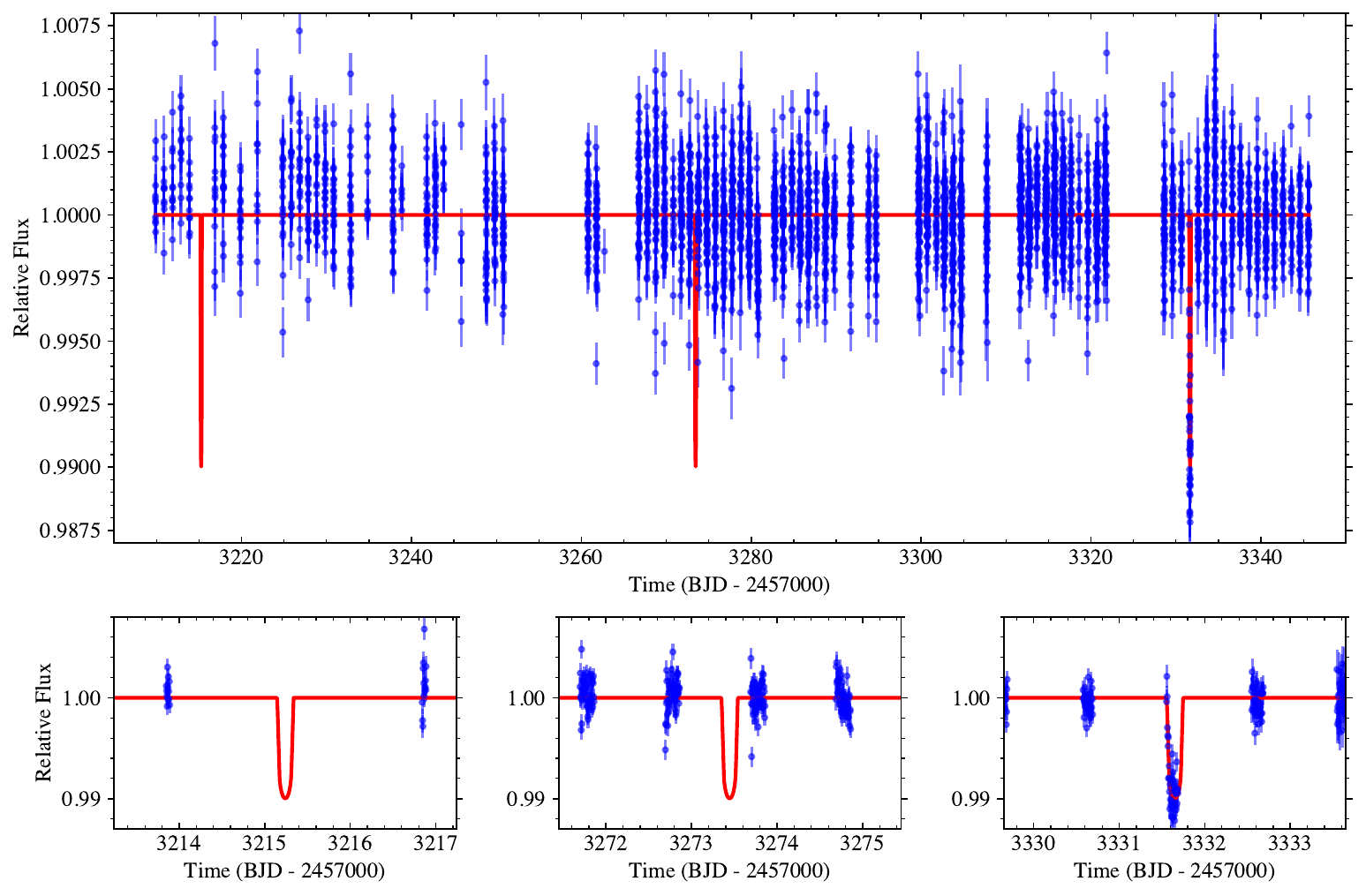}
    \caption{\NGTS\ lightcurves from 2023/09/03 to 2024/02/04 for \starname\ from the \texttt{bsproc} pipeline, normalised and detrended. The transit event is visible in the lower-right panel and was observed on the night of 2024/01/21. For clarity, all data are binned to 5-minute cadence.}
    \label{fig:ngts_sectors}
\end{figure*}

We monitored \starname\ with the NGTS facility for a total of 104 nights. In total this yielded 82,594 photometric data points. For the present analysis, we excluded nights containing fewer than 200 data points, resulting in the removal of 16 nights. The NGTS photometry from all available observing epochs was combined into a single light curve after restricting the sample to exposures conducted under photometric conditions (i.e. no cloud detected). Non-photometric data, including those affected by cloud, were identified through variations and drops in the photometric zero point. For each exposure, we extracted the barycentric time (BJD\_${\rm TDB}$, offset by 2457000), the processed target flux and associated uncertainty, the sky background, and the photometric zero-point. Since observations were made over all moon conditions, the sky background varied greatly over the 104 nights of the campaign. We therefore carried out a simple linear detrend of the photometric flux against the sky background level. We removed 136 photometric data points that we suspected were affected by thin cloud with zero-point magnitudes outside of the the nominal range $-0.06 < {\rm zp} < 0.2$. We removed 102 photometric data points via an iterative $4\sigma$ clipping with three iterations, which were likely due to effects such as cosmic rays and satellites contaminating the photometry of the target or comparison stars. The resulting lightcurve is set out in \autoref{fig:ngts_sectors}. For illustrative purposes we bin the data into 5-min intervals, requiring at least 20 points per bin.


On the night of BJD 2460331.6503 (2024-01-22 03:35 UTC), we detected an ingress event for \starname\ consistent with the transit depth measured by \tess\ in Sector 33. A second transit was later (after 6 consecutive periods from the transit detected with NGTS) detected in \tess\ Sector 87 on 2025/01/05, confirming an orbital period of 58.2\,days. We therefore scheduled NGTS follow-up for the next visible transit opportunity, which occurred on 2025/10/22. For this campaign, we observed simultaneously with six NGTS telescopes using 10\,s exposures. During this night, we again detected a transit ingress, confirming the orbital period of 58.2\, days.

\subsection{TFOP SG1 Follow-up Photometry}
\label{sec:sg1_phot}
In order to monitor the expected transit event on the night of 2025/10/22, we issued an observing alert to the TFOP SG1 team on 2025/10/20. This alert resulted in several observations from SG1 observers around the globe. The data from these observations were reduced with \texttt{AstroImageJ} \citep{Collins2017}. We list each SG1 observation in this sub-section. Reduced data from each SG1 observation is available on the corresponding ExoFOP \citep{Christiansen_2025_ExoFOP} page.\footnote{\url{https://exofop.ipac.caltech.edu/tess/target.php?id=453147896}}

\subsubsection{Acton Sky Portal (0.36 m)}\label{sec:Acton-Sky-Portal (0.36 m)}
The Acton Sky Portal is a private observatory in Massachusetts, USA. The telescope features a Rowe-Ackermann Schmidt Astrograph (RASA 11) from Celestron with a f/11 focal ratio and a 0.36\,m primary mirror. The telescope is equipped with a back-illuminated SBIG Alumna CCD4710 camera, featuring a $1056 \times 1027$ pixel sensor and with 13\,$\mu$m pixel pitch having an image scale of $\sim1\arcsec\,\mathrm{pixel}^{-1}$, resulting in a $17.1\arcmin\times17.1\arcmin$ full field of view \citep{GAPS_PROGRAMME}. An egress event of \planetname\ was observed on 2025/10/22 using the $r'$ filter and an exposure time of 20\,s. An uncontaminated $8''$ target aperture was used for photometry, yielding a 10\,ppt transit event. 

\subsubsection{LCO-CTIO (0.35 m)}\label{sec:LCO-CTIO (0.35 m)}
Observations were obtained with a 0.35-meter telescope of the Las Cumbres Observatory Global Telescope \citep[LCOGT;][]{Brown2013} network, specifically at its Cerro Tololo Inter-American Observatory node in Chile. This instrument is a PlaneWave DeltaRho 350 telescope tube paired with a QHY600 CMOS camera \citep{Harbeck2024_LCOGT_0.35m}. This configuration yields an image scale of $\sim0.74\arcsec\,\mathrm{pixel}^{-1}$ and a $114\arcmin\times72\arcmin$ full field of view. Raw images were calibrated using the standard \texttt{BANZAI} pipeline \citep{McCully2018_LCOGT_BANZAI}. An ingress event of \planetname\ was observed on 2025/10/22 with the $r'$ filter and an exposure time of 60\,s. An uncontaminated $7.4''$ target aperture was used for photometry, yielding a 10\,ppt transit event. 

\subsubsection{ULMT (0.6 m)}\label{sec:ULMT (0.6 m)}
The University of Louisville Manner Telescope (ULMT) is a 0.6\,m Ritchey–Chrétien telescope with an f/8 focal ratio, providing a high-quality, coma-free field of view. The telescope is equipped with a ZWO ASI 6200 CMOS camera, featuring a $9600 \times 6400$ pixel sensor with 3.76\,$\mu$m pixel pitch. This configuration yields a field of view of approximately $25.9\arcmin \times 17.2\arcmin$ with a pixel scale of $\sim 0.16\arcsec\,\mathrm{pixel}^{-1}$, enabling precise time-series photometric observations. An egress event of \planetname\ was observed on 2025/10/22 using the $r'$ filter and with exposure time of 60\,s. An uncontaminated $5''$ target aperture was used for photometry, yielding a 10\,ppt transit event. 

\subsubsection{Lookout Observatory (0.27 m)}\label{sec:Lookout Observatory (0.27 m)}

The Lookout Observatory, a private observatory in Carefree, Arizona, USA,  operates a 0.27\,m primary mirror telescope modified with lens in the secondary, dedicated for wide-field imaging. The telescope is equipped with a ZWO ASI2600 CMOS camera, featuring a $6248 \times 4176$ pixel sensor with 3.76\,$\mu$m pixel pitch. This configuration provides a field of view of approximately $114\arcmin \times 86\arcmin$ with a pixel scale of $\sim 2.9\arcsec\,\mathrm{pixel}^{-1}$ \citep{lookout}. An egress event of \planetname\ was observed on 2025/10/22 using the $r$, $g$, and $b$ filters with 30\,s exposures. Photometry was performed with an uncontaminated $30''$ aperture, yielding a transit depth of 10\,ppt.

\subsubsection{OACC - CAO (0.6 m)}\label{sec:OACC - CAO (0.6 m)}
The Campo Catino Austral Observatory (OACC-CAO) is a facility located in El Sauce Observatory in the Atacama desert, in Chile. The telescope features a 0.6\,m PlaneWave CDK24 Corrected Dall–Kirkham optical design, delivering a wide, flat, and coma-free field of view. The telescope is fitted with a CMOS detector, specifically in the Moravian C3-61000EC PRO camera. This camera employs a large-format sensor (9576 $\times$ 6388) with 3.76 microns pixel pitch, providing a field of view of approximately $32\arcmin \times 21\arcmin$ with a pixel scale of $\sim 0.2\arcsec\,\mathrm{pixel}^{-1}$. The camera is equipped with the $i2$ filter. \planetname\ was observed on the night of 2025/10/22 and an ingress event was detected with an exposure time of 120\,s. An uncontaminated $12''$ target aperture was used for photometry, yielding a 10\,ppt transit event. 

\subsubsection{KeplerCam (1.2 m)}\label{sec:KeplerCam (1.2 m)}
The Fred Lawrence Whipple Observatory (FLWO) in southern Arizona is equipped with a 1.2\,m telescope \citep{kepplercam}. The telescope is fitted with KeplerCam, a conventional CCD imager based on the Fairchild CCD 486 detector. The large-format sensor, comprising $4096 \times 4096$ pixels with a pixel pitch of 15 microns, provides a field of view of approximately $23\arcmin \times 23\arcmin$, making it well suited for high-precision transit photometry. An egress event was detected for the target \planetname\, on the night of 2025/10/22, with the $i'$ filter and an exposure time of 120\,s. An uncontaminated $12''$ target aperture was used for photometry, yielding a 10\,ppt transit event. 

\begin{table*}
    \centering
    \caption{Summary of photometric observations of NGTS-39.}
    \label{tab:phot_summary}
    \begin{tabular}{cccccccccc}
        \toprule
        Telescope &  Night(s) Observed &  $N_\text{images}$ &  Exptime (s) & $N_\text{nights}$\textbf{*} & Filter & Sector & Pipeline  & Comments\\
        \hline
        \tess & 2019/01/08-2019/02/16 & 1072 & 1800 & 22.33 & \tess & 7 &  \texttt{lightcurve}  &  No transit \\
        \tess & 2020/12/18-2021/01/13 & 3485 & 600 & 24.20  & \tess & 33 & \texttt{lightcurve}   & Transit \\
        \tess & 2021/01/14-2021/02/08 & 8961 & 200 & 20.74 & \tess & 71 & \texttt{lightcurve}   & No transit \\
        \tess & 2023/01/18-2023/02/12 & 9577 & 200 & 22.17 & \tess & 72 & \texttt{lightcurve}   & No transit \\
        \NGTS & 2023/09/04-2024/02/04 & 82594  & 10 & 104    & \NGTS & N/A & \bsproc  & Ingress:2024/01/21 \\
        \tess & 2024/12/18-2025/01/14 & 9360 & 200 & 21.67  & \tess & 87 & \texttt{lightcurve}  & Transit \\
        \NGTS\textbf{**}       & 2025/10/22 & 3788 & 10 & 1  & \NGTS & N/A & \bsproc & Ingress\\
        Acton-Sky-Portal      & 2025/10/22 & 762 & 20  & 1  & $r'$  & N/A & \texttt{AstroImageJ}  & Ingress \\
        LCO-CTIO              & 2025/10/22 & 141 & 60  & 1  & $r'$  & N/A & \texttt{AstroImageJ}  & Ingress \\
        ULMT                  & 2025/10/22 & 77  & 60  & 1  & $r'$  & N/A & \texttt{AstroImageJ}  & Egress \\
        Lookout               & 2025/10/22 & 551 & 30  & 1  & $OSC$ & N/A & \texttt{AstroImageJ}  & Egress \\
        OACC-CAO              & 2025/10/22 & 74  & 120 & 1  & $i2$  & N/A & \texttt{AstroImageJ}  & Ingress \\
        KeplerCam             & 2025/10/22 & 305 & 120 & 1  & $i'$  & N/A & \texttt{AstroImageJ}  & Egress \\
        \bottomrule
    \end{tabular}
    \begin{tablenotes}
      \small
      \item \textbf{*} For \tess\ this is calculated by (Exptime (s) $\times N_\text{images})/86400$\,s while for all ground based instruments this is simply the count of nights with observations.
      \item \textbf{**} For \NGTS\ on 2025/10/22, simultaneous observations conducted with 6 telescope units.
    \end{tablenotes}
\end{table*}

\begin{center}
\begin{table}
    \centering
    \caption{A portion of the \NGTS\ photometric data for NGTS-39. The full table is available online.}
    \label{tab:ngtsphot}
    \begin{tabular}{l c c c c }  
    \hline\hline
    Time (BJD   & Normalised flux & Flux uncertainty \\
    -2457000) & & \\
    \hline
3331.5383449076 & 1.008609148 & 0.007050435 \\
3331.5386111111 & 1.011612138 & 0.007044304 \\
3331.5387615738 & 1.006352026 & 0.007027734 \\
3331.5389004629 & 1.011517066 & 0.007022968 \\
3331.5390509260 & 1.012522132 & 0.007017232 \\
3331.5392013890 & 1.018593763 & 0.007018303 \\
3331.5395023148 & 1.006691687 & 0.006990872 \\
3331.5396527778 & 0.998240140 & 0.006981137 \\
3331.5398032409 & 0.988144640 & 0.006964347 \\
    \multicolumn{1}{c}{...} & \multicolumn{1}{c}{...} & \multicolumn{1}{c}{...} \\
    \hline
    \end{tabular}
\end{table}
\end{center}

\section{Spectroscopy} 
\label{sec:rv}

\subsection{Radial Velocity monitoring with \coralie} 
\label{sec:rv_coralie}
RV measurements were taken for \starname\ with the \coralie\ instrument \citep[][]{Queloz2001CORALIE}. \coralie\ is a high-resolution, fiber-fed echelle spectrograph mounted on the 1.2\,m Leonard Euler Swiss Telescope at La Silla Observatory in Chile. It delivers a resolving power of $R \approx 60{,}000$ and operates in the visible wavelength range, yielding an accuracy of approximately 3-5\,ms$^{-1}$ \citep{Segransan_2010, Fontanet_2025}. The effective precision depends on the target properties and observing conditions. In total, we collected 43 RV measurements across two observing epochs: 14 data points between 2023/04/16 and 2024/04/09, and 29 data points between 2024/10/28 and 2026/01/29. We excluded in total 4 data points, due to noisy CCF and high airmass (>1.74). An instrument upgrade occurred between the two epochs, introducing an additional RV offset; consequently, we treat CORALIE14 and CORALIE24 as separate instruments throughout our analysis. CORALIE has demonstrated long-term Doppler stability in planet-search programmes spanning more than two decades for suitable targets \citep{Segransan_2010, Rickman2019}. All observations were acquired with an exposure time of 1500\,s. The spectra were reduced using the standard \coralie\ reduction pipeline (DRS 3.3.12), and RV measurements derived from standard cross-correlation techniques with a numerical
F9 mask. In our data, the median RV uncertainties of the \coralie\ measurements are $23.7\,{\rm m\,s^{-1}}$ for CORALIE14 and $21.5\,{\rm m\,s^{-1}}$ for CORALIE24. These uncertainties are substantially larger than the intrinsic \coralie\ precision. The CORALIE RVs are therefore dominated by photon noise rather than by the instrumental precision floor. The \coralie\ data is set out in \autoref{tab:specdata}. These measurements, together with the \harps\ data show a signal matching with the photometric period and yielding the mass of the planet.

\subsection{Radial Velocity monitoring with \harps} 
\label{sec:rv_harps}
We obtained spectra of \starname\ using \harps\ \citep[][]{Pepe2002HARPS}, the high-resolution, fiber-fed echelle spectrograph mounted on the ESO 3.6-m telescope at La Silla Observatory in Chile. \harps\ provides a resolving power of $R \approx 115{,}000$ across the visible band and is optimised for ultra-stable, high-precision RV measurements. In total, we collected 17 RV measurements by the Warm Jupiter program
(program numbers are : 112.261U.002, 112.261U.003, 116.28XG.001,  116.29B6.001) using exposure time that varied between 1500\,s to 1800\,s for each observation. Three measurements were taken between 2024/04/07 and 2024/04/11, seven between 2025/02/19 and 2025/11/21, and four between 2025/10/16 and 2025/10/26. For our \harps\ data, the median RV uncertainty is $6.3\,{\rm m\,s^{-1}}$. This is larger than the typical HARPS instrumental accuracy of $\sim 1\,{\rm m\,s^{-1}}$. This effective precision depends on the target properties and observing conditions. The \harps\ RV uncertainties are therefore dominated by photon noise rather than by the instrumental precision floor. 

The data were reduced using the \harps\ Data Reduction Software (DRS 3.3.6). Radial velocities were derived via cross-correlation with the F9 mask, providing measurements of the RV, CCF full width at half maximum, bisector span, and associated uncertainties \citep{Baranne1996ELODIE, Pepe2002HARPS}. The \harps\ data is set out in \autoref{tab:specdata}. These data showed a signal consistent with the \coralie\ data indicating a Jupiter-mass exoplanet in a 58\,day orbital around \starname.  The \harps\ spectra of \starname\ were also co-added and used for determining the stellar parameters (see \autoref{sec:Stellar analysis of the system}).

\section{Analysis and Results} 
\label{sec:Results and Analysis}
\subsection{Stellar analysis of the system}
\label{sec:Stellar analysis of the system}
The spectroscopic parameters ($T_{\mathrm{eff}}$, $\log g_{\star}$, [Fe/H]) were estimated using the \texttt{ARES+MOOG} method described in \citealt[][]{Santos-13, Sousa_14, Sousa-21}. For this we used the \texttt{ARES} code\footnote{The last version, \texttt{ARES} v2, can be downloaded at \url{https://github.com/sousasag/ARES}} \citep{Sousa-07, Sousa-15} to measure the equivalent widths (EWs) for the list of \ion{Fe}{i} and \ion{Fe}{ii} lines presented in \citet[][]{Sousa-08}. The spectrum analysed was obtained by co-adding the individual exposures performed by HARPS on the target (co-added S/N \, $\approx$ \, 110 per pixel at 5500\, \AA), such that the uncertainty on the derived parameters is not limited by the spectra S/N, but by the intrinsic precision and accuracy of the method instead (for details see \citet[][]{Figueira_2025}). The best set of spectroscopic parameters was determined by using a minimization process to find the ionization and excitation equilibrium using a grid of Kurucz model atmospheres \citep{Kurucz-93} and the latest version of the radiative transfer code \texttt{MOOG} \citep{Sneden-73}. We also measured $v\sin{i_{\star}}$ from isolated metal lines. Moreover, we derived a more accurate trigonometric surface gravity using Gaia DR3 \citep{Gaia2023GaiaDR3} data following the same procedure as described in \citet[][]{Sousa-21}, which provides a value consistent with the spectroscopic one.

To determine the stellar radius of NGTS-39, we utilised a Markov chain Monte Carlo (MCMC) modified infrared flux method \citep[IRFM --][]{blackwell_1977,schanche_2020}. By building a spectral energy distribution from stellar atmosphere models \citep{castelli_2003} using our spectroscopically derived stellar parameters, we obtained synthetic photometry that was compared to observed broadband fluxes in the following bandpasses: 2MASS $J$, $H$, and $K$, WISE $W1$ and $W2$, and \textit{Gaia} $G$, $G_\mathrm{BP}$, and $G_\mathrm{RP}$ \citep{skrutskie_2006,wright_2010,Gaia2023GaiaDR3}. This enables us to compute the stellar bolometric flux, from which we obtained the effective temperature and stellar radius when combined with the offset-corrected \textit{Gaia} parallax \citep{lindegren_2021}.

We finally derived the stellar mass and age after inputting $T_{\mathrm{eff}}$, [Fe/H], $R_{\star}$, and $v\sin{i_{\star}}$ along with their uncertainties in the isochrone placement routine by \citet{bonfanti2015,bonfanti2016}.
The algorithm interpolates the input set within pre-computed grids of PARSEC\footnote{\textsl{PA}dova and T\textsl{R}ieste \textsl{S}tellar \textsl{E}volutionary \textsl{C}ode: \url{https://stev.oapd.inaf.it/cgi-bin/cmd}} v1.2S \citep{marigo2017} isochrones and tracks, further implementing the gyrochronological relation from \citet{barnes2010} to improve convergence as detailed in \citet{bonfanti2016}.
Following \citet{bonfanti2021}, we conservatively inflated the internal uncertainties of the outcomes to account for the isochrone precision and obtained the results listed in Table~\ref{tab:host_properties}.

\subsection{Global modelling of the NGTS-39 system}
\label{sec:Modelling the NGTS-39 system}
We performed global modelling of the \starname\ system to determine the physical and orbital parameters of the exoplanet \planetname. Our analysis combines space-based photometry from \tess, ground-based photometry from \NGTS\ and the SG1 follow-up network, the RV measurements obtained with \harps\ and \coralie, and the stellar parameters derived from the co-added \harps\ spectra, and stellar parameters from \gaia\ \citep{Gaia2016GaiaMission}. 

The modelling was carried out using the \texttt{allesfitter} python package \citep{allesfitter-code, allesfitter-paper}, which enables the simultaneous analysis of multi-instrument light curves and RV data within a unified Bayesian framework. Transit light curves were modelled using the \texttt{ellc} package \citep{maxted2016ellc}, while parameter inference was performed using the nested sampling algorithm implemented in \texttt{dynesty} \citep{speagle2020dynesty}, following the formalism of \citet{Skilling2004nestedsampling} and \citet{Skilling2006nestedsampling}. The sampling was initialised with 500 live points and used a single bounding strategy to enclose the evolving set of live points. New samples were drawn using a random-walk proposal within the likelihood constraint, providing exploration of the parameter space. The nested sampling run was terminated once the remaining contribution to the Bayesian evidence fell below a tolerance of 1\%. For the spectroscopic data, we used the \texttt{sample\_linear} baseline model for each RV instrument. This accounts for the independent velocity zero-point offsets between HARPS and the two CORALIE datasets, while fitting a common linear RV trend across all three instruments.


We used uniform priors for orbital period $P$, radius ratio ($R_\text{p} / R_\text{*}$), semi-major axis ($(R_\text{*} + R_\text{p}) / a$), epoch of inferior conjuction ($T_{0,\text{p}}$), orbital inclination ($\cos{i_p}$). We use normal priors for the quadratic limb darkening coefficients calculated based on the effective temperature, metallicity and the logarithm surface gravity, $\log {g_{\star}}$ of the host star \citep{limb-dark}. For the treatment of limb darkening, we used the stellar effective temperature, metallicity, and surface gravity, $\log {g_{\star}}$, to estimate the quadratic limb-darkening coefficients of the host star \citep{limb-dark}. This was done using the Limb Darkening Tool Kit \citep[\ldtk;][]{Parviainen2015}, which provides the coefficients in the conventional $u_1,u_2$ parametrisation. These coefficients were then converted into the triangular $q_1,q_2$ parametrisation introduced by \citet{Kipping2013}, and applied in \alles\ as normal priors. We fit the eccentricity and argument of periastron of the system (implemented as the terms $\sqrt{e} \cos{\omega}$ and $\sqrt{e} \sin{\omega}$). We also include a dilution factor for the \tess\ data, as the target star was contaminated by star with number 2 and delta magnitude of 4 within the aperture (see \autoref{fig:ind_lcs}). We detect an additional linear RV trend. We therefore fit for a linear RV trend in the data. Independent zero point offset for each spectrograph are also fitted to account for absolute instrumental offsets between \coralie\ (pre and post upgrade) and \harps. All of the fitted parameters are set out in \autoref{tab:ns_table}. 

\begin{figure*}
    \centering
    \begin{subfigure}{\columnwidth}
        \includegraphics[width=\textwidth]{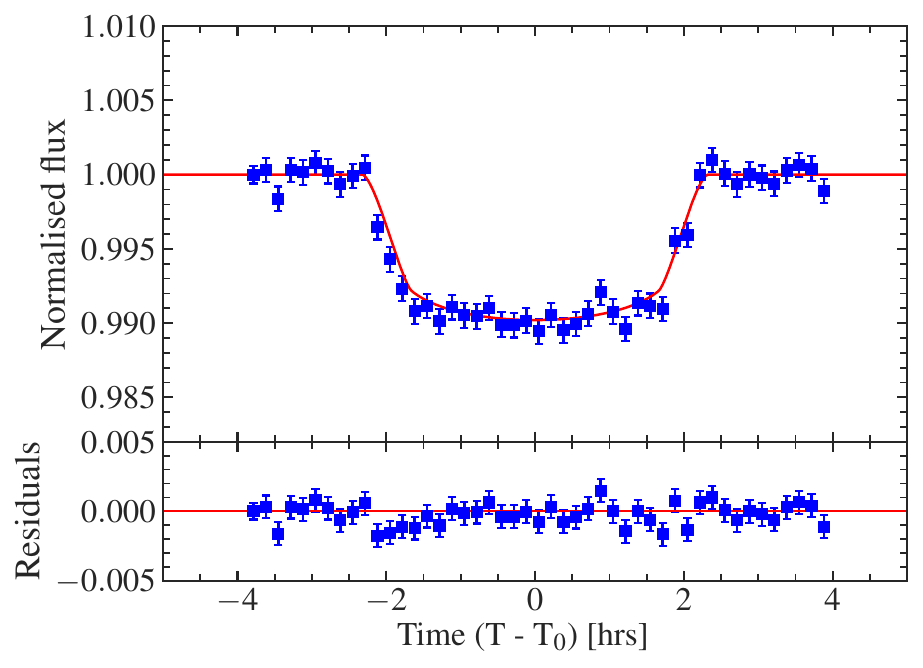}
        \caption{\tess\ Sector 33 transit.}
        \label{fig:tess_33}
    \end{subfigure}
    \begin{subfigure}{\columnwidth}
        \includegraphics[width=\textwidth]{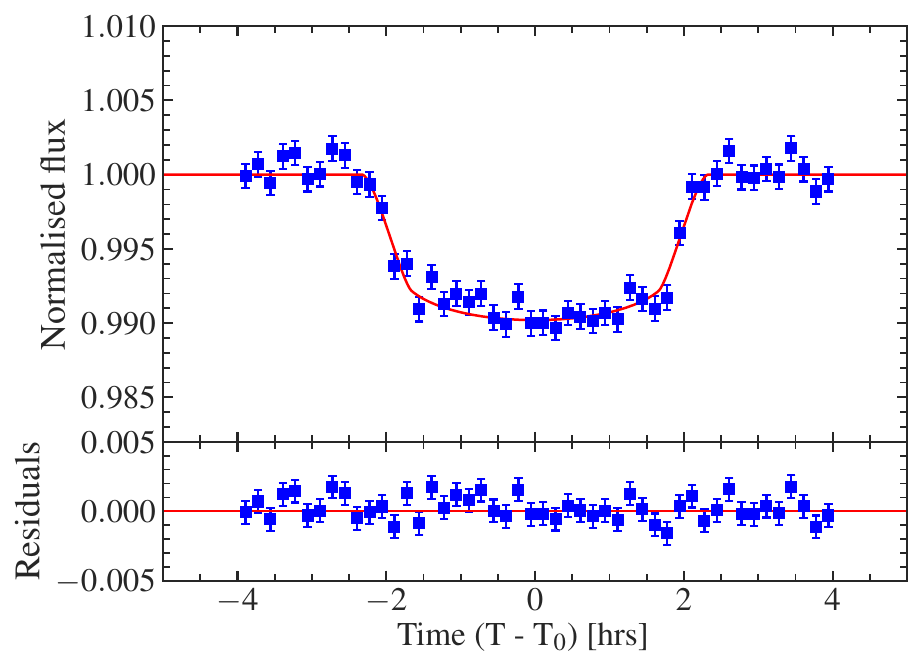}
        \caption{\tess\ Sector 87 transit.}
        \label{fig:tess_87}
    \end{subfigure}
    \caption{Transit lightcurves of \starname\ from \tess. Each panel displays the transit data, depicted by square markers with errors. The data is binned to 10-minutes. The transit model is shown with the solid line. The left panel shows the data from \tess\ Sector 33  centered on BJD=2459225.7606 recorded with 10 minutes cadence and with the \tess\ filter. The right panel shows the data from \tess\ Sector 87 centered on BJD=2460680.8787 recorded with 200\,s cadence and with the \tess\ filter.}
    \label{fig:tess_lc}
\end{figure*}
\begin{figure*}
    \centering
    \begin{subfigure}{\columnwidth}
        \includegraphics[width=\textwidth]{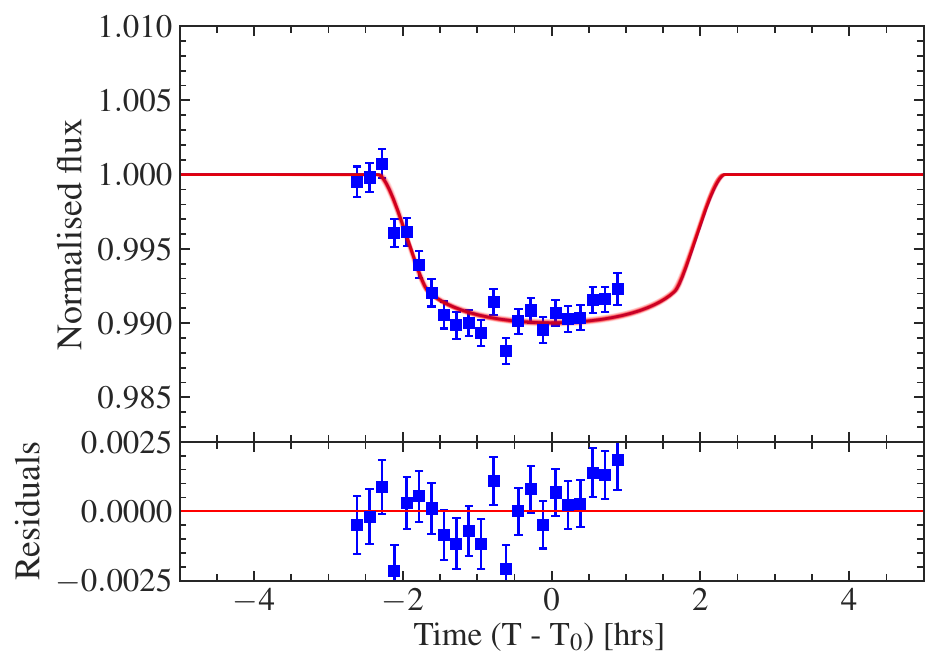}
        \caption{\NGTS\ transit with 1 telescope}
        \label{fig:transits_NGTS_1}
    \end{subfigure}
    \begin{subfigure}{\columnwidth}
        \includegraphics[width=\textwidth]{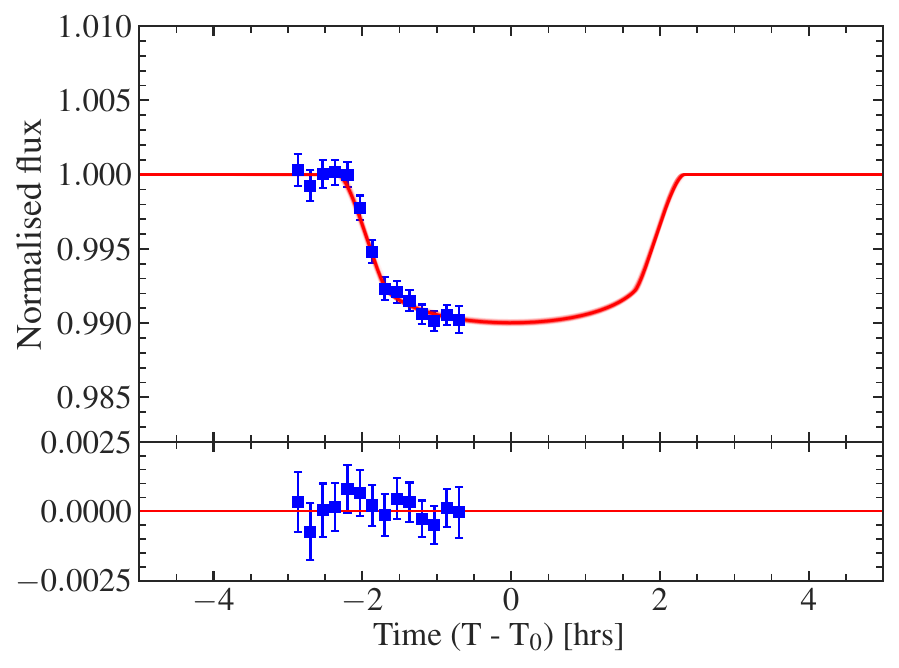}
        \caption{\NGTS\ transit with 6 telescopes}
        \label{fig:transits_NGTS_6}
    \end{subfigure}
    \caption{Transit lightcurves of \starname\ from \NGTS. Each panel displays the transit data, shown as square markers with errors. The data are binned to 10-minutes. The transit model is shown with the solid line. The left panel shows the data from \NGTS\ with 1 telescope unit, with 10\,s exposure using the NGTS filter from the observing night of 2024/01/21 centered on  BJD=2460331.6503. The right panel shows the data from \NGTS\ with 6 telescope units, with 10\,s exposure using the NGTS filter from the observing night of 2025/10/22 centered on  BJD=2460971.9023.}
    \label{fig:ngts_lc}
\end{figure*}
\begin{figure}
	\includegraphics[width=\columnwidth]{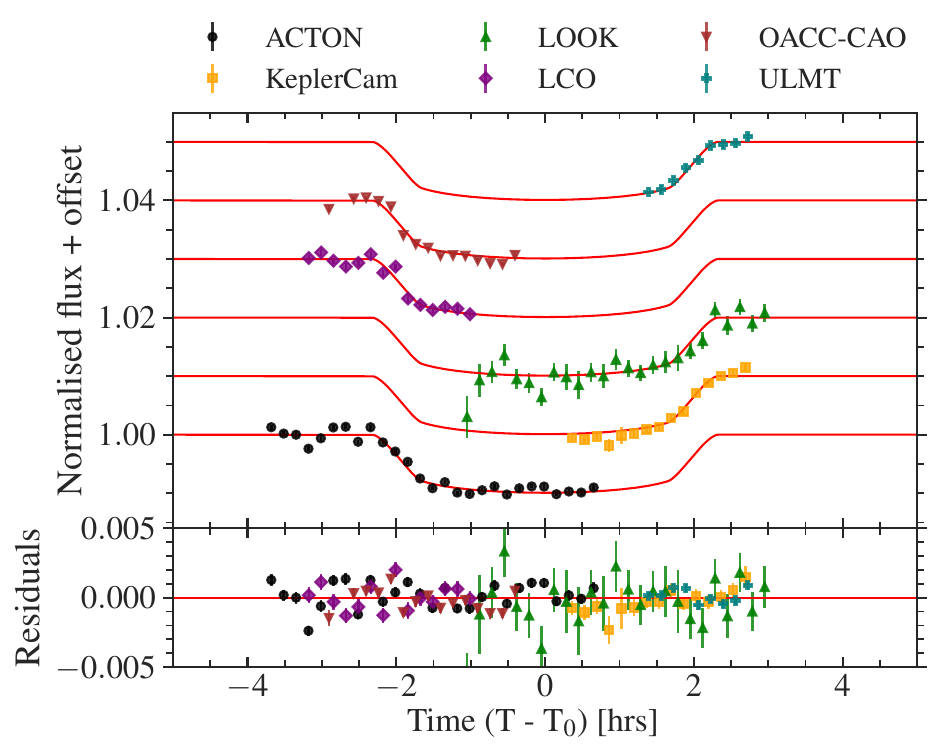}
    \caption{Transit lightcurve of \starname\ normalised to the out-of-transit flux levels taken with LCO-CTIO, ULMT, Lookout, Acton Sky Portal, KeplerCam and OACC-CAO on 2025/10/22 centered on BJD=2460971.9023. The best fit model is shown in red. The data are binned to 10-minutes for clarity. The panel shows the six individual light curves, each vertically offset for clarity.}
    \label{fig:sg1}
\end{figure}
\begin{figure*}
    \centering
    \begin{subfigure}{\columnwidth}
        \includegraphics[width=\textwidth]{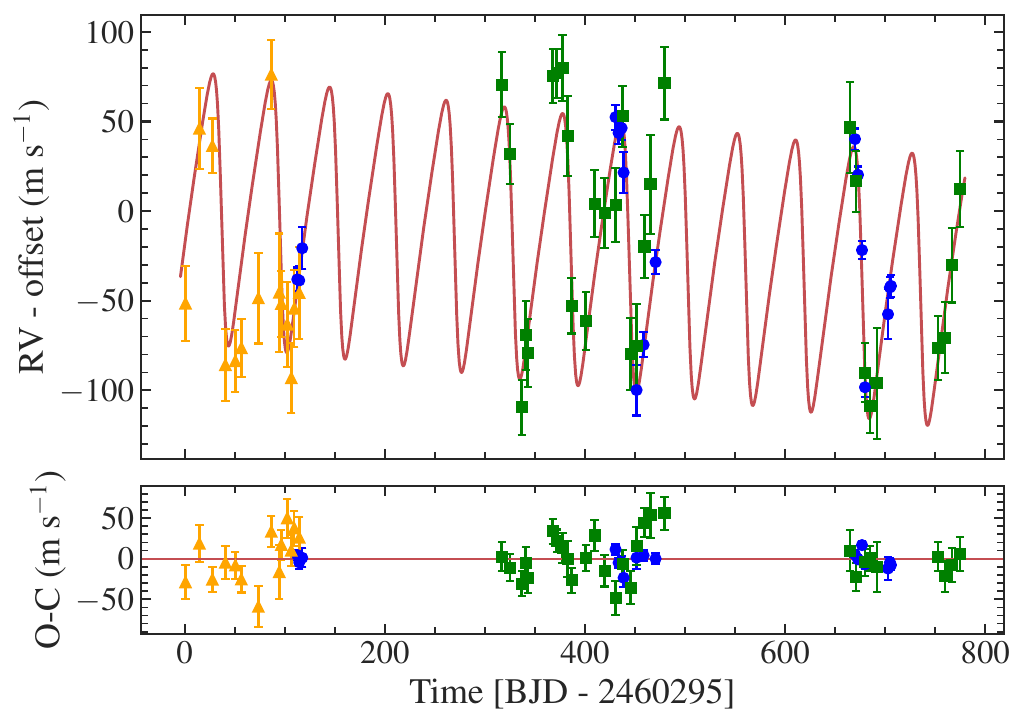}
        \caption{RV versus time of \starname.}
        \label{fig:tic_rv_time}
    \end{subfigure}
    \begin{subfigure}{\columnwidth}
        \includegraphics[width=\textwidth]{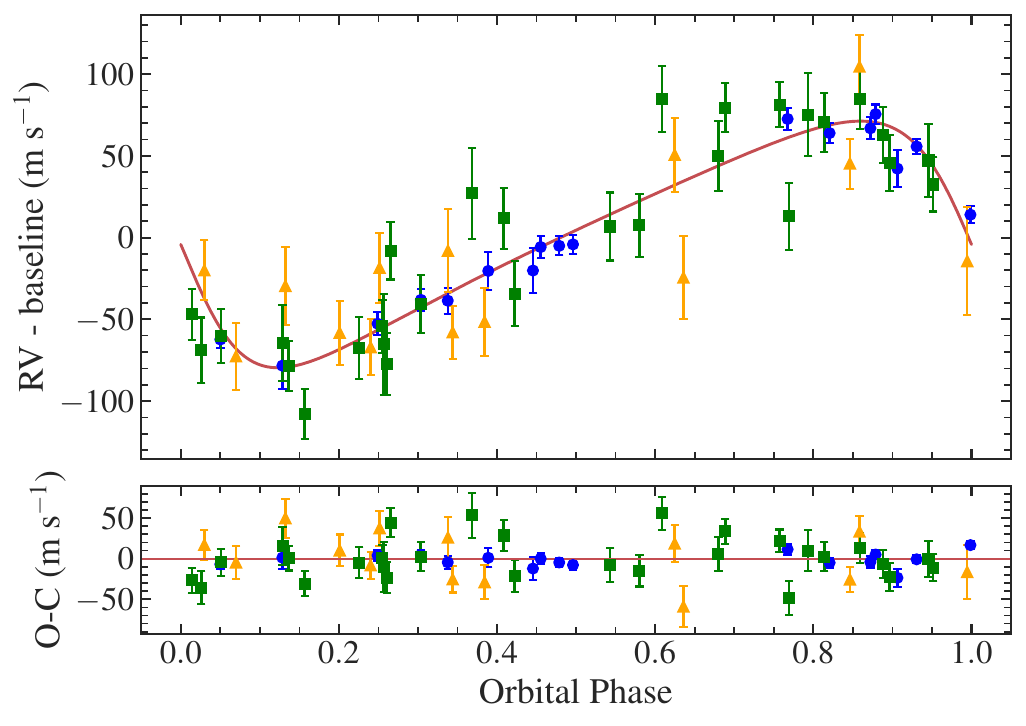}
        \caption{Phase folded radial velocities of \starname.}
        \label{fig:tic_rv}
    \end{subfigure}
    \caption{RV measurements of \starname. \coralie\ data from the first (CORALIE14) and second (CORALIE24) observing cycles are shown as yellow triangles and green squares, respectively, while \harps\ measurements are represented by blue circles. The solid line indicates the best-fit RV model. The left panel presents the radial velocities over time along with the residuals from the fit, and the right panel shows the phase-folded radial velocities with their residuals.}
    \label{fig:rvs}
\end{figure*}

Our results of the final global model are presented in
\autoref{tab:planet_properties}, and we overplot corresponding median posterior model as a solid red line on our photometric and RV data in \autoref{fig:tess_lc}, \autoref{fig:ngts_lc}, \autoref{fig:sg1}, and \autoref{fig:rvs}. The global model derived values of the system are presented in \autoref{tab:ns_derived_table}.

\begin{table}
    \centering
    \caption{NGTS-39\,b properties.}
    \label{tab:planet_properties}
    \begin{tabular}{ccc}
        \toprule
        Parameter & Value\textbf{*} & Source \\
        \hline
        Period\,(days) & \bperiod & \alles \\
        \hline
        \multicolumn{3}{c}{\it Transit Parameters} \\
        \hline
        T$_0$\,(BJD) & \bepoch & \alles \\
        \rpl/\rstar & \brr & \alles \\
        \rstar/a & \bRstarovera & \alles \\
        $b$ & \bbtra & \alles \\
        T$_{14}$\,(hrs) & \bTtratot & \alles \\
        T$_{23}$\,(hrs) & \bTtrafull & \alles \\
        \hline
        \multicolumn{3}{c}{\it RV parameters} \\
        \hline
        $K$\,(\kms) & \bK & \alles \\
        $e$ & \be & \alles \\
        $\omega$\,(degrees) & \bw & \alles \\
        $\dot{\gamma}$\,(m\,s$^{-1}$\,yr$^{-1}$) & $-17.75 \pm 1.86$ & This Work (\S\ref{sec:The NGTS-39 system}) \\
        \hline
        \multicolumn{3}{c}{\it Derived parameters} \\
        \hline
        \rpl\,(\rjup) & \bRcompanionRjup & \alles \\
        \mpl\,(\mjup) & \bMcompanionMjup & \alles \\
        $\rho_\text{p}$\,(\gcc) & \bdensity & \alles \\
        $a$\,(au) & \baAU & \alles \\
        Periapsis & \periastron & $a(1-e)$ \\
        Apoapsis & \apastron & $a(1+e)$ \\
        \teq\,(K) & \bTeq & \alles \\
        $g_\mathrm{b}$ (cgs) & $3040_{-140}^{+150}$ & \alles \\
        \teq\textsubscript{;Apastron}\,(K) & \bTeqAp & This work (\S\ref{sec:The NGTS-39 system}) \\
        \teq\textsubscript{;Periastron}\,(K) & \bTeqPeri & This Work (\S\ref{sec:The NGTS-39 system}) \\        
        \bottomrule
    \end{tabular}
    \begin{tablenotes}
    \small
    \item \textbf{*} Those values represent the posterior median values.
    \end{tablenotes}
\end{table}

\subsection{The NGTS-39 system}
\label{sec:The NGTS-39 system}
\planetname\ is a Jupiter-sized gas giant exoplanet with a radius of \bRcompanionRjup\,\rjup\ and a mass of \bMcompanionMjup\,\mjup, implying a bulk density of \bdensity\,\gcc, which is consistent with a composition dominated by hydrogen and helium, although its density is higher than that of Jupiter and may indicate an enhanced heavy-element content. The planet completes one full orbit every \bperiod\, days with a semi-major axis of \baAU\,AU. Owing to its relatively long orbital period, \planetname\ has a moderate mean equilibrium temperature of $519^{+6}_{-5}$\,K, computed assuming a Bond albedo of 0.3, unit emissivity, and full day-night heat redistribution, placing it in the warm Jupiter regime. The planet exhibits a high orbital eccentricity of \be.

The RV data reveals a long-term trend of 
$\dot{\gamma}$ = $-17.75$\,m\,s$^{-1}$\,yr$^{-1}$, which is evident in \autoref{fig:tic_rv_time}, over a total RV baseline spanning 2.11\,years. This trend indicates the presence of an additional, longer-period companion in the system. To assess whether the trend could instead be caused by stellar activity, for example by long-term magnetic variability, we compared the RV residuals retaining the long-term trend with available activity indicators, namely the FWHM of the cross-correlation function and the H-$\alpha$ index. We find no significant correlation between the RV drift and either diagnostic, supporting the interpretation that the observed long-term RV trend is not due to stellar activity.

\begin{figure}
	\includegraphics[width=\columnwidth]{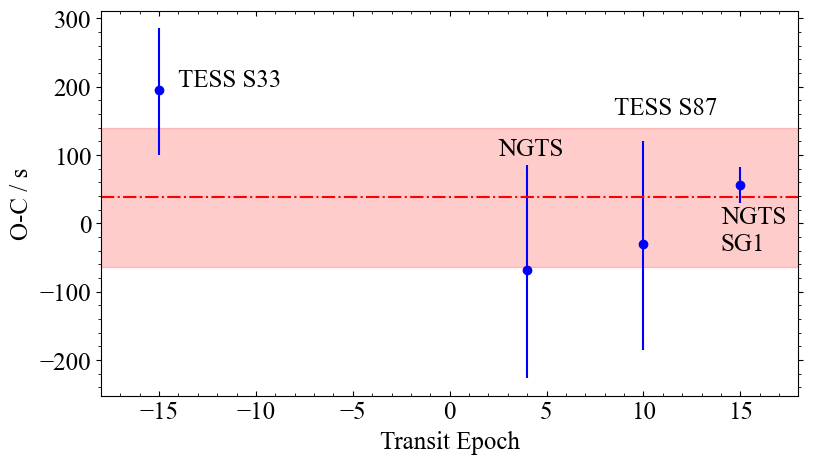}
    \caption{O-C diagram for transit \planetname\ using \tess\ data of Sector 33 and 87 (epochs -15 and 10 respectively), NGTS monitoring (epoch 4) and follow-up data from NGTS and the SG1 network (epoch 15). The mean value of the linear model is shown as a red dashed line and the 1-$\sigma$ RMS with the red shaded area.}
    \label{fig:ttv}
\end{figure}

In addition to the outer longer-period companion, we also searched to see if \starname\ hosts any additional exoplanets. The transit mid-times from the two observed epochs of \tess\ observations, Sectors 33 and 87, the observed ingress from NGTS and the combined lightcurve of NGTS with 6 telescopes and SG1 network were independently fitted to search for deviations from a linear ephemeris, which would indicate transit timing variations (TTVs). A linear fit to the transit times yields an ephemeris consistent with a constant orbital period, with residuals that are consistent with zero within the timing uncertainties. No statistically significant TTV signal is detected across the four transits of \planetname\ - see \autoref{fig:ttv}.

Searches for additional periodic signals in the photometric and RV data did not reveal strong evidence for a second planet. The transit search using all the available photometric data recovered the known signal at $\approx$ 58.2\,d, using the Box Least Square method \citep[BLS;][]{Kovacs2002BLS} consistent with the adapted orbital period found from the joint model. Broader period searches were dominated by alias structure. After masking the transit signal and repeating the analysis on the residual lightcurves, the remaining peaks were weak and lacked coherence expected for a robust additional transiting companion. A complementary period search of the RVs likewise recovered the known planet near 58.2\,d. Additionally, we found a weaker peak at approximately 29.1\,d in the RV periodogram. We note that the absence of corresponding transits (the transits would have been evident in Sector 7 and Sector 71) does not by itself exclude an additional planet, since such a planet could have a non-transiting orbital inclination. We therefore tested whether this signal persists after removing the fitted signal of the known planet. After subtracting the \texttt{allesfitter} posterior-median eccentric RV model for the 58.2\,d planet, together with the fitted RV baselines, we recomputed the periodogram of the RV residuals. The 29.1\,d signal was no longer significant, with a false-alarm probability of 1.0. Since this period is also very close to half of the fitted planet period, $P/2 = 29.102$\,d, we interpret the 29.1\,d peak as the first harmonic of the eccentric 58.2\,d RV signal rather than evidence for an independent planetary signal. We therefore find no statistically evidence for an additional planet in the present photometric and RV data.

\subsection{Sensitivity maps for Companions}
\label{sec:Sensitivity maps for Companions}

\begin{figure}
	\includegraphics[width=\columnwidth]{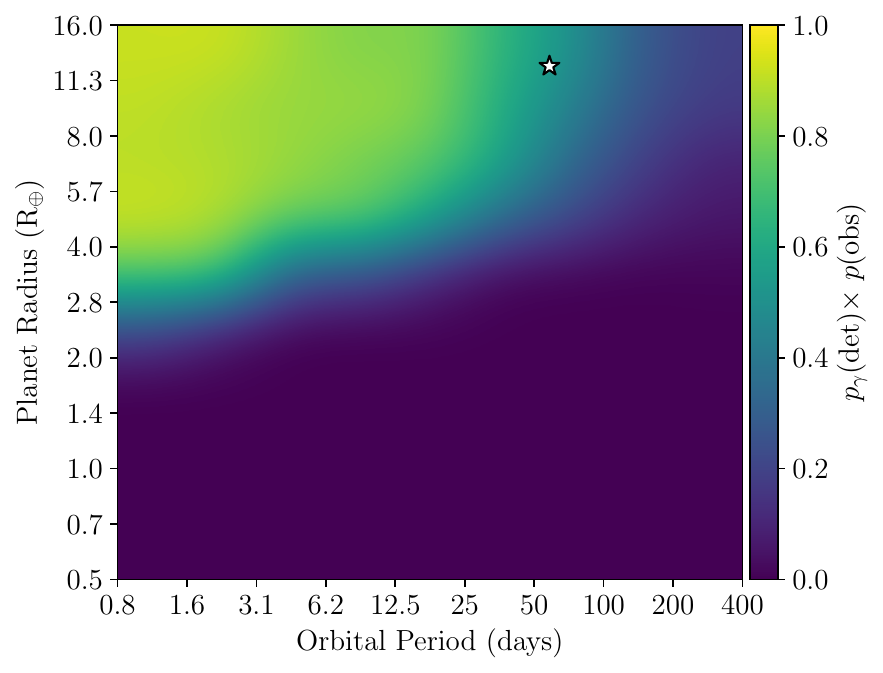}
    \caption{Sensitivity maps from \tiara\ for NGTS-39 in \tess. The target is overplotted as a white star with black borders.}
    \label{fig:tiara}
\end{figure}

In order to make a quantitative assessment of which potential companion planets are ruled out by our data, and which remain allowed, we assessed the detection limits of the \tess\ data for \planetname\ using the Transit Investigation and Recoverability Application \citep[\tiara;][]{Rodel2024}. We used lightcurve timestamps, contamination values, and the measured photometric precision extracted from the lightcurve \texttt{FITS} headers (1-\texttt{CROWDSAP} and \texttt{CDPP2\_0}, respectively) as input for all available sectors. From these inputs, \tiara\ generates synthetic transiting planet parameters and calculates the expected signal-to-noise ratio and detection probability for a range of orbital periods and planetary radii, producing a sensitivity map shown in \autoref{fig:tiara}. The resulting map indicates that \tess\ is most sensitive to large, short-period planets interior to the orbit of \planetname, where multiple transits are more likely to fall within the observing baseline and the transit signal is correspondingly easier to recover. We note that the sensitivity map in \autoref{fig:tiara} does not necessarily reach unity, even for high signal to noise ratio signals, because \tiara\ converts signal to noise ratio into a recovery probability using incomplete gamma functions that asymptote below unity. This recovery probability is then multiplied by the observing-window probability, further limiting the final sensitivity value \citep{Rodel2024}.

In contrast, the sensitivity decreases toward longer orbital periods, reflecting the reduced probability of observing one or more transits and the lower number of available events from which to build detection significance. Long-period ($P > 100$\,d), Jupiter-sized companions are therefore unlikely to be detected by \tess\ around this star. Likewise, planets with radii smaller than Neptune-sized ($\sim 3.8$\,\rearth) are unlikely to be detectable by \tess\ regardless of their orbital period, so the analysis cannot rule out smaller inner or outer companions. Thus, while the existing \tess\ data place useful limits on close-in, large- radius companions, they do not exclude lower-radius planets or giant companions on wider orbits.

\section{Discussion} 
\label{sec:discussion}
\subsection{Parameters of NGTS-39 b}
\label{sec:Space parameter of NGTS-39b}
\begin{figure}
	\includegraphics[width=\columnwidth]{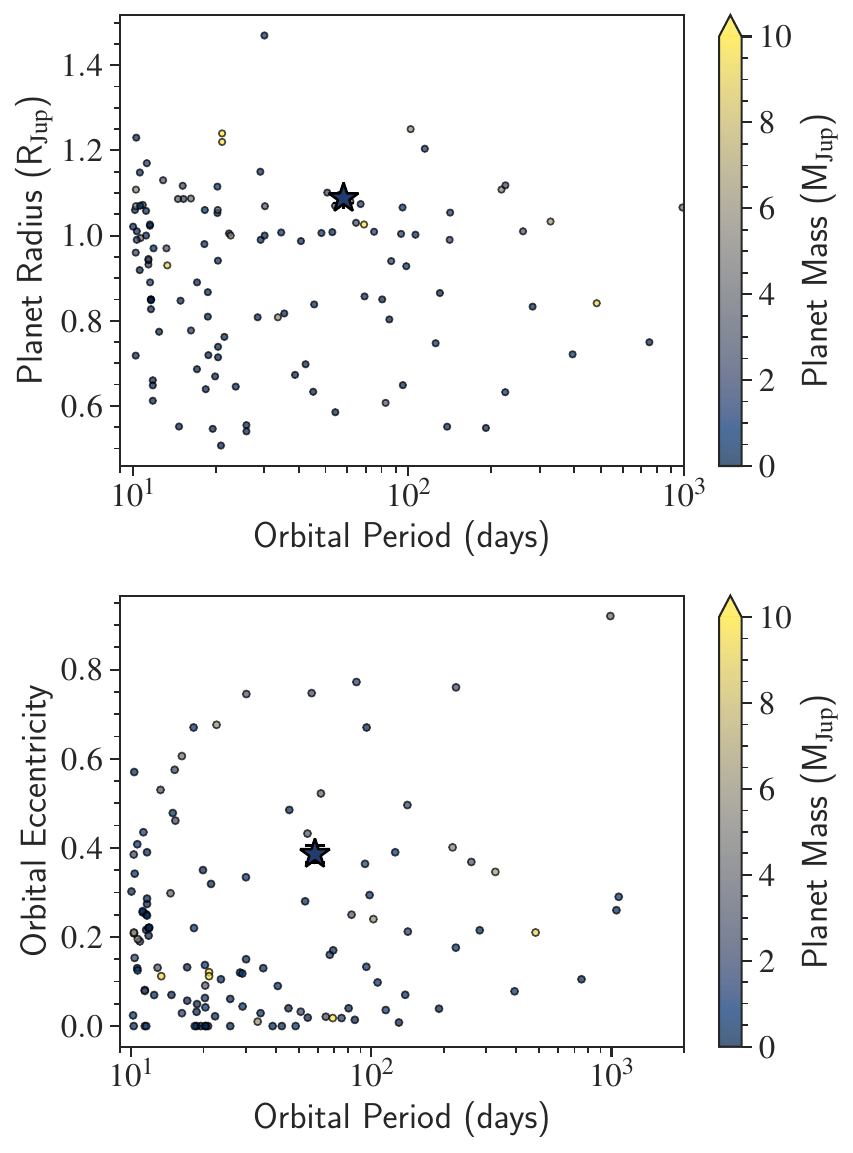}
    \caption{Population diagrams showing planetary radius (top) and orbital eccentricity (bottom) as a function of orbital period for exoplanets with period > 10 days and radius > 0.5 $R_{\mathrm{Jup}}$. \planetname\ is highlighted as a star marker with a black outline and is colour-coded by its mass, consistent with the rest of the planetary sample.}
    \label{fig:period_radius_mass}
\end{figure}

\planetname\ is a long-period exoplanet that contributes to the small but growing population of transiting gas giant planets on wide orbits, as illustrated in the period–radius and period–eccentricity diagrams shown in \autoref{fig:period_radius_mass}. In our sample we included only transiting exoplanet systems with P $> 10$\,days and  M > 0.5 M$_\mathrm{J}$, resulting in 123 exoplanets. In both parameter spaces, \planetname\ occupies a region that is sparsely populated by known exoplanets, highlighting the rarity on detecting of such systems. Its relatively long orbital period places it beyond the bulk of the well-characterised hot Jupiter population, while its high eccentricity further distinguishes it from the predominantly low eccentricity Jupiter-mass planets found at shorter periods.

\subsection{The atmosphere of NGTS-39 b}
\label{sec:The atmosphere of NGTS-39b}
Warm Jupiters are particularly valuable targets for atmospheric studies, as they occupy a transitional regime between hot Jupiters and the colder gas giants of the Solar System, where different chemical and physical processes are expected to dominate \citep{Showman_2020,Fortney2020}.  \planetname\ is therefore an interesting object for atmospheric characterisation because its relatively moderate equilibrium temperature places it in a sparsely explored chemical regime, while its host star is sufficiently hot and bright to make follow-up observations feasible. To place \planetname\ in context, we use the same sample of transiting exoplanets as in \autoref{sec:Space parameter of NGTS-39b}, to examine the relationship between the host star’s effective temperature and the planet’s equilibrium temperature, highlighting the atmospheric regime in which this system resides based on \citet{Fortney2020}.

\begin{figure}
	\includegraphics[width=\columnwidth]{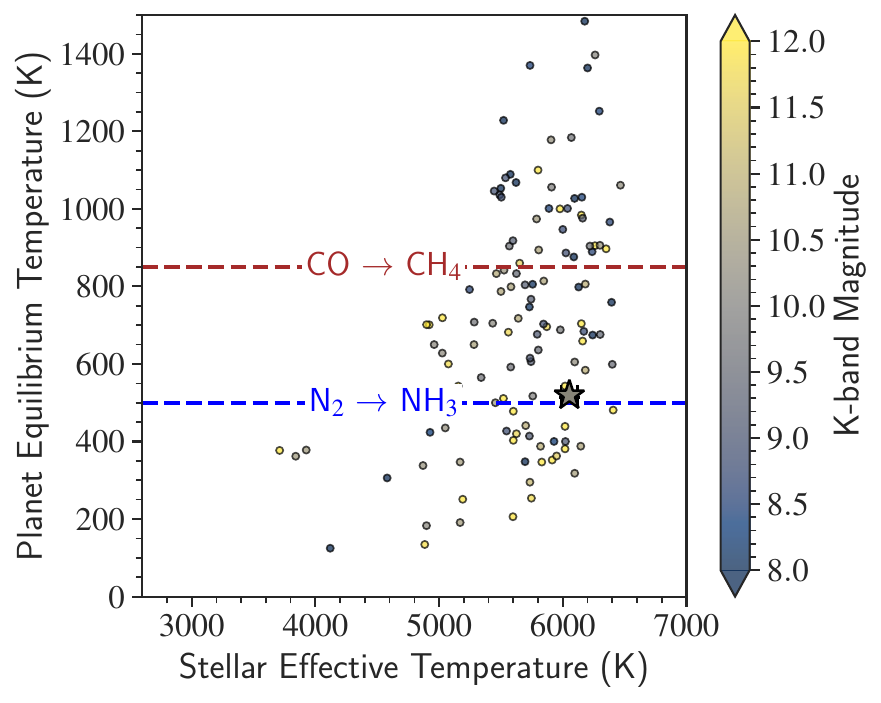}
    \caption{Stellar effective temperature, \teff, is plotted against planetary equilibrium temperature, \teq. \planetname\ is highlighted as a star marker with a black outline and is colour-coded by its orbital eccentricity, consistent with the rest of the planetary sample. The horizontal dashed lines at 850\,K and 500\,K mark the expected CO–CH$_4$ and N$_2$–NH$_3$ chemical transition boundaries, respectively, for a 1\,\mjup\ planet \citep{Fortney2020}.}
    \label{fig:chemistry}
\end{figure}

Our results are presented in \autoref{fig:chemistry}. \planetname\ lies near the transition between
molecular nitrogen (N$_2$) and ammonia (NH$_3$), making it an excellent target for follow-up spectroscopic observations aimed at probing atmospheric chemistry in this temperature regime. We note, however, that the N$_2$-NH$_3$ transition reported by \citet{Fortney2020} is calculated for a 1\,\mjup\ planet, whereas \planetname\ has a mass of approximately 1.5\,\mjup. The location of this transition may therefore shift slightly for \planetname, although we do not expect the effect to be large; if anything, the transition would likely occur at somewhat higher temperature. Additionally, we estimate the planetary temperature variations based on the periastron and apastron distances, taking into account the planet’s high orbital eccentricity. We found a periastron temperature of \bTeqPeri,K and an apastron temperature of \bTeqAp\,K. These values suggest that \planetname\ lies between the CO–CH$_4$ and N$_2$–NH$_3$ chemical transition boundaries at apastron, while at periastron it falls below the N$_2$–NH$_3$ transition.



\subsection{An outer companion}
\label{sec:an_outer_companion_senario}

Since the current RV observations likely cover only a limited fraction of such an outer orbit, the signal appears as an approximately linear acceleration rather than a fully resolved Keplerian signal. Continued RV monitoring is required to detect any curvature in the trend and to constrain the orbital period, eccentricity, and mass of the outer companion.

The Gaia RUWE value of 0.793 indicates a single-star astrometric solution, with no evidence for a significant astrometric signal in the Gaia DR3 data \citep{Castro-Ginard_2024}. This disfavours a stellar-companion capable of producing a large astrometric signal. However, it does not rule out a lower-mass companion, such as a massive planet or brown dwarf, nor a companion on an orbit with a period significantly longer than the 34-month Gaia DR3 observing baseline relevant to the RUWE metric \citep{Guerriero_2026}.


\section{Conclusion}\label{sec:conclusion}
In this work, we report the detection and characterisation of \planetname, a long-period warm Jupiter. A transit was initially observed by \tess\ in Sector 33. The NGTS follow-up group marked it as a monotransit and initiated an observing campaign for follow-up photometry of \starname\ with NGTS. An ingress event of \planetname\ was observed and with another transit in Sector 87 the orbital period was constrained.  Ground-based follow-up photometry from \NGTS\ was further obtained to better demonstrate the orbital period, with additional photometric observations from the SG1 network further refining the planetary parameters. We also acquired spectroscopic RV measurements with \harps\ and \coralie\ to confirm the planetary nature of the signal and measure the planetary mass and orbital properties.

A detailed stellar analysis was performed for \starname, which is found to be a main-sequence F9 star. \planetname\ is a warm Jupiter with a mass of \bMcompanionMjup\,\mjup\ and a radius of \bRcompanionRjup\,\rjup. It completes one orbit every \bperiod\ days, has an equilibrium temperature of 519\,K (assuming a Bond albedo of 0.3), and exhibits a relatively high orbital eccentricity of \be. 

This work highlights the importance of systematic follow-up of single-transit events detected by \tess, which are likely to reveal a substantial and previously under-explored population of long-period exoplanets. Such planets provide crucial constraints on planetary formation and migration pathways, particularly in the regime between hot and cold Jupiters. Expanding this sample will help bridge current observational gaps and improve our understanding of the diversity of giant exoplanet architectures. \planetname\ is also a valuable target for spectroscopic follow-up being close to the boundary where ammonia is expected to become visible. 

Additional RV measurements over a longer time baseline, together with future astrometric constraints, will be required to confirm the nature of the outer companion in the \starname\ system.

\section*{Acknowledgements}
This project was conducted as part of a UK Science and Technology Facilities Council (STFC) Industrial CASE (Cooperative Awards in Science and Technology) PhD studentship. IA is the STFC funded PhD student, and gratefully acknowledges the support STFC under the CASE Industry scheme ST/W005077/1 (Project title: Precision Photometry with the new generation of fast readout Scientific CMOS Camera).
Based on data collected under the NGTS project at the ESO La Silla Paranal Observatory. The NGTS facility is operated by the consortium institutes with support from the UK Science and Technology Facilities Council (STFC) under projects ST/M001962/1, ST/S002642/1 and ST/W003163/1. Based on observations collected at the European Southern Observatory under ESO programmes 112.261U.002, 112.261U.003, 116.28XG.001, and 116.29B6.001. This paper includes data collected by the \tess\ mission. Funding for the \tess\ mission is provided by the NASA Explorer Program. This work has made use of data from the European Space Agency (ESA) mission
{\it Gaia} (\url{https://www.cosmos.esa.int/gaia}), processed by the {\it Gaia}
Data Processing and Analysis Consortium (DPAC,
\url{https://www.cosmos.esa.int/web/gaia/dpac/consortium}). Funding for the DPAC
has been provided by national institutions, in particular the institutions
participating in the {\it Gaia} Multilateral Agreement.
This publication makes use of data products from the Two Micron All Sky Survey, which is a joint project of the University of Massachusetts and the Infrared Processing and Analysis Center/California Institute of Technology, funded by the National Aeronautics and Space Administration and the National Science Foundation. 
This work makes use of observations from the Las Cumbres Observatory global telescope network. This paper is based on observations made with the Las Cumbres Observatory's education network telescopes that were upgraded through generous support from the Gordon and Betty Moore Foundation. This paper is based on observations made with observatory time provided to Boyce Research Initiatives and Education Foundation by the Las Cumbres Observatory through its Global Sky Partners program. 
MPB gratefully acknowledges support from UK Research and Innovation (UKRI) under the UK government’s Horizon Europe funding guarantee for an ERC starting grant [grant number EP/Z000890/1]. FEN is supported by the UKRI (Grants ST/X001121/1, EP/X027562/1). JAE acknowledges support through the European Space Agency (ESA) Research Fellowship Programme in Space Science. PF acknowledges financial support from the Severo Ochoa grant CEX2021-001131-S funded by MCIN/AEI/10.13039/501100011033 and by the European Union (ERC, THIRSTEE, 101164189). Views and opinions expressed are however those of the author(s) only and do not necessarily reflect those of the European Union or the European Research Council. Neither the European Union nor the granting authority can be held responsible for them. FH gratefully acknowledges the support of the NGTS consortium despite no longer working full-time in astronomical research. ML, SU, MH, HC, FB acknowledge support of the Swiss National Science Foundation under grant number PCEFP2\_194576. The contribution of ML, SU, MH, HC, FB has been carried out within the framework of the NCCR PlanetS supported by the Swiss National Science Foundation under grants 51NF40\_182901 and 51NF40\_205606. X.D acknowledges the support from the the Swiss National Science Foundation under the grant SPECTRE (No 200021\_215200).This publication has been made possible by Spanish grants PID2021-125627OB-C31 and PID2024-158486OB-C31 funded by MCIU/AEI/10.13039/501100011033 and by “ERDF A way of making Europe”,  by the programme Unidad de Excelencia María de Maeztu CEX2020-001058-M financed by MCIN/AEI/10.13039/501100011033 and by the MaX-CSIC Excellence Award MaX4-SOMMA-ICE, by the Generalitat de Catalunya/CERCA programme, and by the European Research Council (ERC) under the European Union’s Horizon Europe programme (ERC Advanced Grant SPOTLESS; no. 101140786). TR is supported by an STFC studentship. S.G.S. acknowledge support from FCT through FCT contract nr. CEECIND/00826/2018 and POPH/FSE (EC). C.A.W. would like to acknowledge support from the UK Science and Technology Facilities Council (STFC, grant number ST/X00094X/1). The work of HPO has been carried out within the framework of the NCCR PlanetS supported by the Swiss National Science Foundation under grants 51NF40\_182901 and 51NF40\_205606. T.G.W. acknowledges support from the University of Warwick and UKSA.

\section*{Conflict of Interest}
The authors declare no conflict of interest.
\section*{Data Availability}
The \NGTS\ photometry data are available through the \href{https://archive.eso.org/cms.html}{ESO Science Archive Facility} and from the VizieR archive server hosted by the Universit\'{e} de Strasbourg.\footnote{cdsarc.u-strasbg.fr}
The {\it TESS} data can be accessed through the MAST (Mikulski Archive for Space Telescopes) portal at \url{https://mast.stsci.edu/portal/Mashup/Clients/Mast/Portal.html}. Photometry data from Acton Sky Portal, LCO-CTIO, ULMT, Lookout Observatory, OACC-CAO and KeplerCam are accessible via the ExoFOP-TESS archive at \url{https://exofop.ipac.caltech.edu/tess/target.php?id=453147896}. Reduced \harps\ spectra, derived measurements of radial velocities will be available from the VizieR archive server hosted by the Universit\'{e} de Strasbourg. The \harps\ spectra were obtained under the Warm Jupiters program, ESO programme IDs: 112.261U.002, 112.261U.003, 116.28XG.001, 116.29B6.001.

\section*{Affiliations}
$^{1}$ Centre for Exoplanets and Habitability, University of Warwick, Gibbet Hill Road, Coventry CV4 7AL, UK\\
$^{2}$ Department of Physics, University of Warwick, Gibbet Hill Road, Coventry CV4 7AL, UK\\
$^{3}$ Centre for Space Domain Awareness, University of Warwick, Gibbet Hill Road, Coventry CV4 7AL, UK\\
$^{4}$ Leiden Observatory, Leiden University, P.O. Box 9513, 2300 RA Leiden, The Netherlands\\
$^{5}$ Astrophysics Research Centre, School of Mathematics and Physics, Queen’s University Belfast, Belfast, BT7 1NN, UK\\
$^{6}$ Astronomy Unit, Queen Mary University of London, Mile End Road, London E1 4NS, UK\\
$^{7}$ Acton Sky Portal (private observatory), Acton, MA, USA\\
$^{8}$ Space Research Institute, Austrian Academy of Sciences, Schmiedlstrasse 6, A-8042 Graz, Austria\\
$^{9}$ Observatoire de Genève, Université de Genève, 51 chemin Pegasi, 1290 Sauverny, Switzerland\\
$^{10}$ School of Physics and Astronomy, University of Leicester, Leicester LE1 7RH, UK\\
$^{11}$ European Space Agency (ESA), European Space Research and Technology Centre (ESTEC), Keplerlaan 1, 2201 AZ Noordwijk, The Netherlands\\
$^{12}$ Instituto de Astrof\'{i}sica de Andaluc\'{i}a-CSIC, Glorieta de la Astronom\'{i}a s/n, E-18008 Granada, Spain\\
$^{13}$ Rugby School, Lawrence Sheriff St, Rugby, Warwickshire, CV22 5EH, UK\\
$^{14}$ Campo Catino Astronomical Observatory, Regione Lazio,Guarcino (FR), 03010 Italy \\
$^{15}$ EPFL Laboratoire d’Astrophysique BSP 320, Cubotron, Route de la Sorge CH, 1015 Lausanne Switzerland \\
$^{16}$ Institute of Space Sciences (ICE, CSIC), Carrer de Can Magrans S/N, Campus UAB, Cerdanyola del Vallès, E-08193, Spain \\
$^{17}$ Institut d’Estudis Espacials de Catalunya (IEEC), 08860 Castelldefels (Barcelona), Spain \\
$^{18}$ Instituto de Astrofisica e Ciencias do Espaco, Universidade do Porto, CAUP, Rua das Estrelas, 4150-762 Porto, Portugal \\
$^{19}$ Departamento de Fisica e Astronomia, Faculdade de Ciencias, Universidade do Porto, Rua do Campo Alegre, 4169-007 Porto, Portugal\\
$^{20}$Center for Astrophysics \textbar \ Harvard \& Smithsonian, 60 Garden Street, Cambridge, MA 02138, USA\\
$^{21}$Instituto de Estudios Astrofísicos, Facultad de Ingeniería y Ciencias, Universidad Diego Portales, Av. Ejército Libertador 441, Santiago, Chile\\
$^{22}$Centro de Excelencia en Astrofísica y Tecnologías Afines (CATA), Camino El Observatorio 1515, Las Condes, Santiago, Chile\\
$^{23}$Obserwatorium Astronomiczne Niedźwiady, Szubin, Poland\\
$^{24}$Boyce Research Initiatives and Education Foundation, San Diego, CA, USA\\
$^{25}$Physikalisches Institut, Universität Bern, Gesellschaftsstrasse 6, 3012 Bern, Switzerland\\
$^{26}$Department of Astronautical Engineering, United States Air Force Academy, CO 80840, USA\\
$^{27}$Univ. Grenoble Alpes, CNRS, IPAG, 38000 Grenoble, France\\



\bibliographystyle{mnras}
\bibliography{refs} 




\appendix

\section{RV dataset}

\begin{center}
\begin{table*}
    \centering
    \caption{Spectroscopic data for \starname.}
    \label{tab:specdata}
    \begin{tabular}{c c c c c c c c c c}
    \hline
    \hline
    Instrument & Time (BJD & RV & RV error & FWHM & Bisector & Contrast & H-$\alpha$ & Ca\,II H K & Na D \\
    & -2400000) & ($\rm m\,s^{-1}$) & ($\rm m\,s^{-1}$) & ($\rm m\,s^{-1}$) & ($\rm m\,s^{-1}$) & & & & \\
    \hline
\coralie\,-14 & 60295.80997534 & 42122.52 & 20.82 & 9426.90 & -107.30 & 43.15 & 0.186224 & 0.245857 & 0.341326 \\
\coralie\,-14 & 60309.80438275 & 42220.16 & 22.41 & 9332.33 & 54.11 & 43.64 & 0.203109 & 0.033177 & 0.358816 \\
\coralie\,-14 & 60322.70379500 & 42210.49 & 15.09 & 9383.14 & -52.96 & 43.22 & 0.194693 & 0.124007 & 0.357653 \\
\coralie\,-14 & 60335.71434432 & 42088.20 & 20.14 & 9291.79 & 28.43 & 42.57 & 0.198576 & 0.011834 & 0.365156 \\
\coralie\,-14 & 60345.62997957 & 42090.41 & 16.94 & 9374.09 & -25.19 & 43.39 & 0.183313 & 0.193392 & 0.345591 \\
\coralie\,-14 & 60351.66770422 & 42097.69 & 16.08 & 9333.39 & -8.98 & 43.61 & 0.191872 & 0.184797 & 0.358601 \\
\coralie\,-14 & 60368.65611608 & 42125.52 & 25.20 & 9309.97 & 3.17 & 42.35 & 0.160463 & 0.237732 & 0.356916 \\
\coralie\,-14 & 60381.61307386 & 42250.29 & 19.13 & 9345.37 & -48.46 & 43.01 & 0.182704 & 0.201045 & 0.382469 \\
\coralie\,-14 & 60389.54133268 & 42128.63 & 32.85 & 9329.29 & -79.17 & 41.81 & 0.159824 & 0.199310 & 0.364466 \\ 
\coralie\,-14 & 60391.58859461 & 42122.35 & 18.16 & 9334.37 & 25.24 & 43.05 & 0.190052 & 0.156512 & 0.347210 \\
\coralie\,-14 & 60397.55231894 & 42110.82 & 23.67 & 9408.83 & 49.91 & 42.73 & 0.191502 & 0.049140 & 0.351448 \\
\coralie\,-14 & 60401.55403324 & 42080.88 & 19.54 & 9266.32 & -63.38 & 42.86 & 0.175115 & 0.212246 & 0.345227 \\
\coralie\,-14 & 60404.48954468 & 42119.63 & 21.31 & 9187.39 & -49.41 & 43.26 & 0.160620 & 0.107313 & 0.338301 \\
\coralie\,-14 & 60409.52064332 & 42128.42 & 25.34 & 9348.32 & -20.34 & 43.22 & 0.182833 & 0.138028 & 0.365810 \\
\coralie\,-24 & 60611.83969783 & 42216.72 & 17.84 & 9339.57 & -24.45 & 43.26 & 0.198008 & 0.167301 & 0.365139 \\
\coralie\,-24 & 60619.84957792 & 42178.04 & 16.37 & 9359.11 & -32.96 & 43.50 & 0.184750 & 0.199590 & 0.354664 \\
\coralie\,-24 & 60631.79977263 & 42036.72 & 15.07 & 9261.17 & -20.02 & 42.96 & 0.176178 & 0.140397 & 0.361714 \\
\coralie\,-24 & 60635.80691002 & 42076.78 & 18.94 & 9310.96 & -9.49 & 43.51 & 0.182663 & 0.100758 & 0.352635 \\
\coralie\,-24 & 60637.83314197 & 42066.74 & 18.79 & 9327.81 & -30.63 & 43.57 & 0.195988 & 0.134071 & 0.351991 \\
\coralie\,-24 & 60662.75537131 & 42221.50 & 14.93 & 9297.96 & -52.07 & 42.54 & 0.186954 & 0.156429 & 0.373003 \\
\coralie\,-24 & 60666.75489358 & 42223.01 & 13.63 & 9308.58 & -70.23 & 43.70 & 0.187874 & 0.079251 & 0.368996 \\
\coralie\,-24 & 60672.67503791 & 42225.85 & 18.33 & 9347.91 & 20.81 & 43.64 & 0.179017 & 0.201322 & 0.345683 \\
\coralie\,-24 & 60677.71792751 & 42187.95 & 22.08 & 9258.96 & -29.21 & 43.53 & 0.187195 & 0.273395 & 0.354393 \\
\coralie\,-24 & 60681.69018643 & 42093.45 & 15.40 & 9364.65 & -55.30 & 43.68 & 0.188171 & 0.119931 & 0.354403 \\
\coralie\,-24 & 60695.72248834 & 42084.92 & 15.83 & 9473.43 & 9.45 & 43.28 & 0.166405 & 0.088115 & 0.361517 \\ 
\coralie\,-24 & 60704.65294822 & 42150.33 & 18.49 & 9364.99 & -13.09 & 43.15 & 0.200977 & 0.243559 & 0.362203 \\
\coralie\,-24 & 60714.63558010 & 42145.20 & 19.23 & 9413.86 & -16.62 & 43.42 & 0.188258 & 0.064294 & 0.348071 \\
\coralie\,-24 & 60725.65281709 & 42149.76 & 20.49 & 9347.55 & -79.72 & 43.29 & 0.179946 & 0.159661 & 0.351143 \\
\coralie\,-24 & 60732.58889236 & 42199.03 & 17.00 & 9309.83 & -39.47 & 43.59 & 0.182394 & 0.087075 & 0.359829 \\
\coralie\,-24 & 60740.59269182 & 42066.56 & 20.00 & 9360.40 & -30.31 & 43.20 & 0.196144 & 0.189669 & 0.349143 \\
\coralie\,-24 & 60746.57986256 & 42070.76 & 23.22 & 9415.97 & -22.77 & 42.51 & 0.196159 & 0.110846 & 0.360828 \\
\coralie\,-24 & 60754.52654126 & 42126.40 & 17.42 & 9317.42 & -36.01 & 43.22 & 0.183655 & 0.182711 & 0.350245 \\
\coralie\,-24 & 60760.52503366 & 42161.09 & 27.64 & 9451.71 & -102.94 & 43.19 & 0.209401 & 0.155811 & 0.350728 \\ 
\coralie\,-24 & 60774.50584590 & 42217.57 & 20.14 & 9304.39 & -26.90 & 42.84 & 0.212061 & 0.147045 & 0.358480 \\
\coralie\,-24 & 60959.87659787 & 42192.80 & 25.42 & 9351.20 & -95.18 & 42.53 & 0.199527 & 0.184917 & 0.403718 \\
\coralie\,-24 & 60965.87765547 & 42162.82 & 16.95 & 9270.30 & -95.66 & 43.58 & 0.178159 & 0.129352 & 0.393471 \\
\coralie\,-24 & 60974.85339112 & 42056.03 & 16.38 & 9418.28 & -43.59 & 42.98 & 0.189350 & 0.192727 & 0.370193 \\
\coralie\,-24 & 60979.84728090 & 42037.55 & 15.11 & 9349.95 & 14.49 & 43.12 & 0.185702 & 0.109920 & 0.363239 \\
\coralie\,-24 & 60986.84233893 & 42050.12 & 30.90 & 9176.93 & -47.38 & 43.11 & 0.200588 & 0.161048 & 0.339410 \\
\coralie\,-24 & 61047.76949862 & 42069.92 & 17.65 & 9341.74 & -7.98 & 43.93 & 0.202864 & 0.145602 & 0.340958 \\
\coralie\,-24 & 61054.69166230 & 42075.52 & 19.57 & 9457.04 & -41.79 & 44.20 & 0.179439 & 0.131281 & 0.352853 \\
\coralie\,-24 & 61061.71616546 & 42115.98 & 20.65 & 9355.73 & -32.19 & 44.10 & 0.186088 & 0.033686 & 0.339726 \\
\coralie\,-24 & 61069.68088692 & 42158.41 & 21.11 & 9424.26 & 13.76 & 43.94 & 0.176873 & 0.080444 & 0.355464 \\
\harps & 60407.53390987 & 41980.80 & 6.15 & 8757.43 & -7.08 & 53.74 & 0.203691 & 0.107009 & 0.340540 \\
\harps & 60409.50864809 & 41980.20 & 7.51 & 8733.59 & -10.50 & 53.75 & 0.215756 & 0.084198 & 0.339823 \\
\harps & 60412.48923116 & 41998.21 & 11.37 & 8760.99 & -21.93 & 53.41 & 0.228377 & 0.045618 & 0.336937 \\
\harps & 60725.54961517 & 42071.35 & 6.52 & 8723.64 & -34.40 & 54.14 & 0.222461 & 0.053107 & 0.339989 \\
\harps & 60728.63808268 & 42062.51 & 5.58 & 8695.46 & -37.23 & 54.22 & 0.206239 & 0.068017 & 0.339346 \\
\harps & 60731.64197968 & 42065.24 & 6.35 & 8739.46 & -14.50 & 54.15 & 0.201473 & 0.059961 & 0.348973 \\
\harps & 60733.63286984 & 42040.44 & 10.98 & 8778.35 & 3.42 & 53.75 & 0.224288 & 0.027084 & 0.358363 \\
\harps & 60746.56373218 & 41919.05 & 13.98 & 8694.36 & -33.72 & 52.92 & 0.257298 & 0.054322 & 0.361998 \\
\harps & 60753.57069333 & 41944.29 & 6.56 & 8746.92 & -64.82 & 53.82 & 0.209537 & 0.043312 & 0.348260 \\
\harps & 60765.57675288 & 41990.42 & 6.30 & 8736.96 & -8.24 & 53.97 & 0.212411 & 0.086548 & 0.351648 \\
\harps & 60964.84191604 & 42059.11 & 5.34 & 8705.22 & -15.77 & 54.03 & 0.217466 & 0.105728 & 0.374481 \\
\harps & 60967.86233549 & 42039.13 & 3.97 & 8742.01 & -55.68 & 53.91 & 0.205740 & 0.100019 & 0.356626 \\
\harps & 60971.82641747 & 41997.14 & 4.55 & 8735.92 & -42.63 & 53.91 & 0.209265 & 0.085147 & 0.365751 \\
\harps & 60974.81643393 & 41920.59 & 4.55 & 8759.67 & -30.39 & 53.82 & 0.210837 & 0.075404 & 0.368474 \\
\harps & 60997.82760728 & 41961.33 & 13.63 & 8685.85 & 30.97 & 53.96 & 0.245599 & -0.062214 & 0.363174 \\
\harps & 60999.76036696 & 41976.32 & 5.27 & 8712.58 & -32.39 & 53.95 & 0.216451 & 0.095447 & 0.354793 \\
\harps & 61000.77443000 & 41977.14 & 5.43 & 8707.87 & -32.62 & 54.09 & 0.214688 & 0.072879 & 0.348836 \\
    \hline
    \end{tabular}
\end{table*}
\end{center}

\section{\alles\ results tables}

\begin{table*}
	\centering
	\caption{\alles\ global fitted values and priors}
	\label{tab:ns_table}
	\begin{tabular}{c c c c}
		\toprule
		Parameter & Prior & Fitted value & units\\
		\hline
		$R_b / R_\star$&$\mathcal{U}\left(0.09,0.1\right)$&\brr&\\
		$(R_\star + R_b) / a_b$&$\mathcal{U}\left(0.01,0.02\right)$&\brsuma&\\
		$\cos{i_b}$&$\mathcal{U}\left(0.002,0.8\right)$&\bcosi&\\
		$T_{0;b}$&$\mathcal{U}\left(2460680.8,2460680.9\right)$&\bepoch&bjd\\
		$P_b$&$\mathcal{U}\left(58,59\right)$&\bperiod&d\\
		$\sqrt{e_b} \cos{\omega_b}$&$\mathcal{U}\left(-1,1\right)$&\bfc&\\
		$\sqrt{e_b} \sin{\omega_b}$&$\mathcal{U}\left(-1,1\right)$&\bfs&\\
		$K_b$&$\mathcal{U}\left(0.060,0.090\right)$&\bK&$\mathrm{km\,s^{-1}}$\\
        $\dot{\gamma}$ &$\mathcal{U}\left(-0.05,0.05\right)$&\baselineslopervHARPS&$\mathrm{km\,s^{-1}\,yr^{-1}}$\\
		$\gamma$ CORALIE14&$\mathcal{U}\left(41.50,42.50\right)$&\baselineoffsetrvCORALIEone&$\mathrm{km\,s^{-1}}$\\
		$\gamma$ CORALIE24&$\mathcal{U}\left(41.50,42.50\right)$&\baselineoffsetrvCORALIEtwo&$\mathrm{km\,s^{-1}}$\\
		$\gamma$ HARPS&$\mathcal{U}\left(41.50,42.50\right)$&\baselineoffsetrvHARPS&$\mathrm{km\,s^{-1}}$\\
        $D_\mathrm{0; TESS}$&$\mathcal{U}\left(0,0.03\right)$&\bdepthtrdilTESS&\\
        $q_{1; \mathrm{NGTS_1}}$&$\mathcal{N}\left(0.3443,0.0105\right)$&\hostldcqoneNGTSone&\\
        $q_{2; \mathrm{NGTS_1}}$&$\mathcal{N}\left(0.3820,0.0817\right)$&\hostldcqtwoNGTSone&\\
		$q_{1; \mathrm{NGTS_6}}$&$\mathcal{N}\left(0.3443,0.0105\right)$&\hostldcqoneNGTSsix&\\
		$q_{2; \mathrm{NGTS_6}}$&$\mathcal{N}\left(0.3820,0.0817\right)$&\hostldcqtwoNGTSsix&\\
        $q_{1; \mathrm{sg1}}$&$\mathcal{N}\left(0.2879,0.0084\right)$&\hostldcqonesgone&\\
		$q_{2; \mathrm{sg1}}$&$\mathcal{N}\left(0.3734,0.0785\right)$&\hostldcqtwosgone&\\
		$q_{1; \mathrm{TESS}}$&$\mathcal{N}\left(0.2880,0.0085\right)$&\hostldcqoneTESS&\\
		$q_{2; \mathrm{TESS}}$&$\mathcal{N}\left(0.3734,0.0785\right)$&\hostldcqtwoTESS&\\
        
		\bottomrule
	\end{tabular}
\end{table*}

\begin{table}
    \centering
    \caption{\alles\ global model derived values}
    \label{tab:ns_derived_table}
    \begin{tabular}{c c}
        \toprule
        Parameter & Value\\ 
        \hline 
        $R_\star/a_\mathrm{b}$ & $0.01751_{-0.00026}^{+0.00028}$\\ 
        $a_\mathrm{b}/R_\star$ & $57.10\pm0.91$\\ 
        $R_\mathrm{b}/a_\mathrm{b}$ & $0.001690\pm0.000031$\\ 
        $R_\mathrm{b}$ ($\mathrm{R_{\oplus}}$) & $12.20\pm0.13$\\ 
        $R_\mathrm{b}$ ($\mathrm{R_{jup}}$) & $1.088\pm0.012$\\ 
        $a_\mathrm{b}$ ($\mathrm{R_{\odot}}$) & $66.2\pm1.2$\\ 
        $a_\mathrm{b}$ (AU) & $0.3077\pm0.0055$\\ 
        Inclination b; $i_\mathrm{b}$ (deg) & $88.961\pm0.041$\\ 
        $e_\mathrm{b}$ & $0.386\pm0.019$\\ 
        $w_\mathrm{b}$ (deg) & $98.3\pm3.1$\\ 
        $q_\mathrm{b}$ & $0.001212_{-0.000040}^{+0.000042}$\\ 
        $M_\mathrm{b}$ ($\mathrm{M_{\oplus}}$) & $466\pm26$\\ 
        $M_\mathrm{b}$ ($\mathrm{M_{jup}}$) & $1.467\pm0.081$\\ 
        $M_\mathrm{b}$ ($\mathrm{M_{\odot}}$) & $0.001400\pm0.000078$\\ 
        $b_\mathrm{tra;b}$ & $0.639_{-0.023}^{+0.020}$\\ 
        $T_\mathrm{tot;b}$ (h) & $4.639\pm0.035$\\ 
        $T_\mathrm{full;b}$ (h) & $3.325\pm0.046$\\ 
        $\rho_\mathrm{\star;b}$ (cgs) & $1.039\pm0.049$\\ 
        $\rho_\mathrm{b}$ (cgs) & $1.411\pm0.092$\\ 
        $g_\mathrm{b}$ (cgs) & $3040_{-140}^{+150}$\\ 
        $T_\mathrm{eq;b}$ (K) & $519.1_{-5.3}^{+6.1}$\\ 
        $\delta_\mathrm{tr; undil; b; TESS}$ (ppt) & $9.91\pm0.17$\\ 
        $\delta_\mathrm{tr; dil; b; TESS}$ (ppt) & $9.785_{-0.14}^{+0.098}$\\ 
        $\delta_\mathrm{tr; undil; b; NGTS_6}$ (ppt) & $9.976\pm0.075$\\ 
        $\delta_\mathrm{tr; dil; b; NGTS_6}$ (ppt) & $9.976\pm0.075$\\ 
        $\delta_\mathrm{tr; undil; b; NGTS_1}$ (ppt) & $9.972\pm0.076$\\ 
        $\delta_\mathrm{tr; dil; b; NGTS_1}$ (ppt) & $9.972\pm0.076$\\ 
        $\delta_\mathrm{tr; undil; b; sg1}$ (ppt) & $9.915_{-0.074}^{+0.068}$\\ 
        $\delta_\mathrm{tr; dil; b; sg1}$ (ppt) & $9.915_{-0.074}^{+0.068}$\\ 
        $u_\mathrm{1; TESS}$ & $0.363\pm0.080$\\ 
        $u_\mathrm{2; TESS}$ & $0.172\pm0.080$\\ 
        $u_\mathrm{1; NGTS_1}$ & $0.331\pm0.082$\\ 
        $u_\mathrm{2; NGTS_1}$ & $0.252\pm0.082$\\ 
        $u_\mathrm{1; sg1}$ & $0.390\pm0.075$\\ 
        $u_\mathrm{2; sg1}$ & $0.148\pm0.075$\\ 
        $rho_\mathrm{\star; combined}$ (cgs) & $1.039\pm0.049$\\ 
        \bottomrule
    \end{tabular}
\end{table}

\bsp	
\label{lastpage}
\end{document}